\documentclass[aps,prb,showpacs]{revtex4}
\usepackage{amsmath,amssymb}
\usepackage{graphicx}
\usepackage{dcolumn}
\usepackage{bm}
\usepackage{color}
\newcommand{\mbb}{\mathbb}
\newcommand{\mc}{\mathcal}

\newcommand{\tet}{\texttt}

\begin{document}
\title{
Finite-temperature Coulomb Excitations in
Extrinsic Dirac Structures
}
\author{Andrii Iurov}
\email{aiurov@unm.edu}
\affiliation{
Center for High Technology Materials, University of New Mexico,
1313 Goddard SE, Albuquerque, NM, 87106, USA
}
\author{Godfrey Gumbs}
\affiliation{
Department of Physics and Astronomy, Hunter College of the City
University of New York, 695 Park Avenue, New York, NY 10065, USA
}
\affiliation{
Donostia International Physics Center (DIPC),
P de Manuel Lardizabal, 4, 20018 San Sebastian, Basque Country, Spain
}
\author{Danhong Huang}
\affiliation{
Air Force Research Laboratory, Space Vehicles Directorate,
Kirtland Air Force Base, NM 87117, USA
}
\affiliation{
	Center for High Technology Materials, University of New Mexico,
	1313 Goddard SE, Albuquerque, NM, 87106, USA
}
\author{Ganesh Balakrishnan}
\affiliation{
        Center for High Technology Materials, University of New Mexico,
        1313 Goddard se, Albuquerque, NM, 87106, USA
}

\date{\today}

\begin{abstract}
We have derived algebraic, analytic expressions for the chemical potential without any restriction on temperature
for all types of doped, or extrinsic, gapped Dirac cone materials including gapped graphene, silicene, germanene 
and single-layer transition metal dichalcogenides. As an important intermediate step of our derivations, 
we have established a reliable piecewise-linear model for calculating the density-of-states in molybdenum disulfide, 
showing good agreement with previously obtained numerical results. For the spin- and valley-resolved 
band structure, we obtain an additional decrease of the chemical potential due to thermally induced doping 
of the upper subband at finite temperature. It has been demonstrated that since the symmetry between the electron and hole
states in $MoS_2$ is broken, the chemical potential could cross the zero-energy level at sufficiently high 
temperature. These results allow us to investigate the collective properties, polarizability, plasmons and their 
damping.  Emphasis is placed on low temperatures, when initial electron doping plays a crucial role.
We clearly demonstrated the contribution of the initial doping to the finite-temperature collective properties 
of the considered materials.
\end{abstract}


\maketitle

\section{Introduction}
\label{s1}

Despite the fact that the microscopic properties of various low-dimensional materials have been 
meticulously examined over a fairly long period of time, \,\cite{andostern, fstern} only successful fabrication of graphene 
in 2004 \,\cite{gr1, gr2} stimulated an intriguingly new  research effort devoted to the study of atomically thin 
two-dimensional (2D) materials. In particular, it was by virtue of its unique, yet unexpected massless Dirac
electronic properties that led to high mobility (200,000 cm$^2$/V$\cdot$s) and ballistic transport properties. 
\,\cite{gr3, neto2, gr5} In the corners of the Brillouin zone, referred to as $K$ and $K'$ points, 
there is no energy band gap and the dispersions represent a linear Dirac cone structure. Due to the
existence of such an energy spectrum, opening a sufficiently large  and tunable energy gap in 
graphene has become an important issue in order to enable electron confinement. Researchers tried to 
achieve this by adjoining  a variety of  insulating 
substrates \,\cite{gap1, gap2, gap3, gap4} or even  expos graphene to circularly-polarized radiation.
\,\cite{kibis}  In finite-width nanoribbons, their energy band structure and gap are modified by the type
of insulating `cousin' that is introduced.\cite{nrb2, nrb3, hanGap}

\par
\medskip
\par

In order to create a truly tunable band gap, one must use a material with large spin-orbit coupling
or a buckled structure. In this regard, silicene, a 2D silicon structure was deemed a good 
candidate. Single-monolayer $Si$ possesses a buckled structure  simply because of the larger ionic size of silicon
compared to carbon. This results in a large spin–orbit band gap $1.55\,meV$ and the possibility to modify
its energy spectrum by applying an external perpendicular electric field \,\cite{ezawa, SilMain, Dru6}. These properties 
make it  display an experimentally realizable Kane–Mele type of quantum spin Hall effect, or a topological insulator state, 
because of the existence of time-reversal symmetry. \,\cite{KaneMele, ezawaprl} Unlike graphene, the band structure of 
silicene and its nanoribbons \cite{drummond, Dru4, Dru5}  directly depends on spin and valley indices
which give lead to plenty of nanoelectronic, valleytronic and spintronic applications.

\par
\medskip
\par

Germanene, the most recently discovered and fabricated member of atomically thin buckled 2D honeycomb lattices, 
\,\cite{G1zhang, G2acun, G11li, G12davila, G13bampoulis, G14derivaz} demonstrates substantially larger
Fermi velocities and a band gap of $20 - 90 \, meV$. Grown by molecular beam epitaxy \,\cite{dacapito} and 
investigated with x-ray absorption spectroscopy, $Ge$ layers demonstrated satisfactory agreement between
the experimentally obtained and theoretically predicted results for its inter-atomic distance.

\par
\medskip 
\par 
Another important class of innovative 2D materials is represented by direct-gap transition metal 
dichalchogenides, or TMDC's. Its chemical makeup consists of a transition metal atom \tet{M}, such as molybdenum
or tungsten, and two identical chalcogens \tet{C}, i.e., sulfur, selenium or tellurium. Schematically, TMDC's are 
described as \tet{MC}$_2$. In our consideration, we are mostly going to focus on $MoS_2$, as
their most studied representative. This material exhibits a semiconductor energy band structure with a very large 
direct gap $1.78\,eV$, in contrast to its bulk states with indirect gap $1.3\,eV$, and substantial spin-orbit
coupling. \,\cite{ganatraACS} Strictly speaking, $MoS_2$ is not a Dirac material since the mass terms play a crucial 
role in its energy dispersions, however its low-energy Hamiltonian contains a $t_0 a_0 \, \Sigma \cdot {\bf k}$ term,
corresponding to the linear Dirac cone dispersion.

\par
\medskip
\par
An effective two-band continuum model and lattice Hamiltonian, \cite{Habib} based on the tight-binding model,
accounts for the hybridization of the \tet{d} orbitals in $Mo$ and the \tet{p} orbitals of sulfur atoms.
It gives an adequate description of its low-energy band structure and predicts large spin splitting. \cite{LargeSpin}
Due to the breaking of inversion symmetry and spin-orbit coupling, spin and valley physics is observed
in all group-IV dichalcogenides, including $MoS_2$. \cite{xiao-prl}  The low-energy states of such systems
are no longer massive Dirac fermions since there is a a difference between electron and hole masses as well as 
trigonal warping effects. \,\cite{korm} Strain engineering, used to tune optical and electronic properties
of conventional semiconductors, has also been  applied to molybdenum disulfide, and its modified band structure
has been theoretically calculated. \,\cite{Habibstrain} These unique electronic properties of a single-layer 
$MoS_2$  were later used to create high-performance transistors operating at room temperature. \,\cite{transistor} 
These electronic models and effective Hamiltonian have been widely used to investigate the collective properties 
of TMDC's \,\cite{Scholz1} and their influence on the gap transition. \,\cite{wangmany} In optoelectronics, the 
band structure, spin and valley properties of molybdenum disulfide could be successfully controlled by 
off-resonant dressing field. \,\cite{kibisall}

\par
\medskip
\par

Current many-body and quantum field theory methods in condensed matter physics \cite{Gui, Gbook} have provided 
helpful ways to understand the electronic and transport properties of low-dimensional solids, including 
diverse bucked honeycomb materials. \,\cite{TabNicPRL,TabNicACDC,TabNicMagneto}  In most 
of these theories, we find the \textit{dynamical polarization function}, or polarizability, to be the mainstay, 
fundamental quantity, describing the screening of an external potential by interacting electrons. \,\cite{AndoS, 
pavlo1, SDSscr, pavlo2, Scholz1}    
Also, the dynamic polarization function plays a key role in calculating the plasmon excitations, due to 
the charge  density oscillations, which occur in metals and semiconductors. Specifically, 
the plasmon dispersion relation  along with their lifetimes  have been theoretically investigated for a wide range of 2D 
Dirac systems. \,\cite{wunsch, SDS07, SDSft, stauber1, stauber2, pavlo1,stauber17} 
The interest in graphene plasmons is due in part to the fact that  these excitations
have no classical counterpart. \,\cite{purelyQ} There has also been a considerable experimental effort 
for investigating graphene plasmons,  gate-tuning, infrared nano-imaging and 
confinement \,\cite{ExpPlas1, ExpPlas2, ExpPlas3, ExpPlas4, ExpPlas5}  
Graphene plasmonic resonances and instability at various wavelengths could be used in photodetectors 
in the Terahertz range. \,\cite{spiler} All these techniques could be successfully applied to the recently fabricated
materials, discussed in the present work.

 \par
 \medskip
 \par 
 
 Plasmonic applications have been widely based on \textit{nanoscale hybrid systems}, in which graphene 
plasmons are coupled to a surface plasmon excitations in metals. Technology has now gone a long way in combining 
graphene with prefabricated plasmonic nanoarrays and metamaterials in order to produce plasmonics-based tunable
hybrid optical devices. \,\cite{Pnano} Therefore, accurate knowledge of plasmon mode dispersions in graphene interfaced 
with metallic substrates is crucial. Graphene-metal contacts are important components for all such devices. 
Consequently, exploration of plasmon modes at these metallic interfaces is a mandatory step toward fabricating the devices. 
High-resolution electron energy loss spectroscopy (EELS)  has been employed to investigate those excitations at the surface of 
$Bi_2 Se_3$ to disclose the interplay between surface and Dirac plasmons in topological insulators \,\cite{poli1}. Plasmons, 
their behavior, dispersions, quenching and environmental effects have been thoroughly studied in epitaxial
graphene, in air-exposed graphene-$Ru$ contacts, Graphene on $Pt_3Ni$ $(1 1 1)$, graphene grown on $Cu$ $(1 1 1)$ 
foils. \,\cite{poli2, poli3, poli4, poli171, poli172}

\par
\medskip
\par

In all cases under consideration, we need to distinguish between \textit{extrinsic}, or  a sample initially 
doped at $T=0$, and intrinsic materials, with zero Fermi energy and completely empty conduction band. 
In the latter case, both the plasmon excitations and the electrical conductivity are completely suppressed 
at zero temperature due to the absence of free charge carriers. However, at  finite temperature, the
conduction band would receive thermally induced doping in both cases. \,\cite{SDSLi, patel} In graphene, 
with zero energy band gap, this density is enhanced as $n \backsimeq T^2$ and the plasmon dispersion 
behaves like $\Omega_p^2 \backsimeq q T$.

\par
\medskip
\par
The properties of intrinsic finite-temperature plasmon excitations have been systematically examined 
for various materials including silicene. \,\cite{W} In contrast, extrinsic or 
doped structure at finite temperature is associated with a difficulty to obtain a reliable 
and accurate  value for temperature-dependent chemical potential $\mu(T)$. 
Generally, it is known that $\mu(T)$ is decreased as the temperature is increased, and its value could be found 
based on carrier density conservation \,\cite{SDSS, SDSLi} In this work, our main objective is to obtain a set 
of non-integral, trancendental equations for a wide class of Dirac gapped materials with linear 
density-of-states (DOS), i.e., gapped graphene, silicene, germanene and transition metal dichalcogenides at arbitrary temperature.

\par
However, the range of our considered temperatures is limited by validity of linear or gapped Dirac cone 
approximation for the energies, which recieve noticeable doping at those temperatures. Certain deviations
start to build up at about $0.5 \, eV$, \cite{LeniP, PP} leading to various effect on the plasmons, such 
as anisotropy, splitting and existence of additional acoustic plasmon branch.  \cite{SilkinP, DesP} Such energies
are extremely far away from our range $k_B T \backsim E_F^{(0)}$. In all our calculations, the energy is measured in
the units of a typical Fermi energy  $E_F^{(0)} = 5.22\, meV$, corresponding to electron density $n^{(0)} = 1.0 \cdot 10^{15}
\,m^{-2}$.

\par
\medskip
\par
Once the chemical potential is known, one can obtain the finite-temperature dynamical polarization 
function by Eq.~\eqref{Pi0T}, which is a key component for all relevant many-body calculations. 
These include  optical absorption, electronic transport, plasmon  excitations as well as electron 
exchange and correlation energies. \,\cite{ourjpcm17} Here, we pay close attention to 
finite-temperature plasmons, demonstrating how much initial doping contributes to each branch location at intermediate
temperatures. Once the temperature becomes very high, $k_B T \gg E_F$, thermally-induced doping dominates and the 
contribution from the initial Fermi energy fades away.

\par
\medskip
\par 

The rest of the paper is organized as follows. First, we derive the implicit analytic equations for the 
finite-temperature chemical potential for all types of Dirac structures with linear DOS in Sec.\  
\ref{s2}. In Sec.~\ref{s3}, we calculate the dynamic polarization function which includes the single
particle excitation mode frequencies. The single-particle modes combine with a charge cloud to 
produce weakly interacting quasiparticles that vibrate collectively at the characteristic plasma frequencies.   
Emphasis has been placed on rather simple cases of gapped graphene and silicene at small, but finite 
temperatures, when the zero temperature carrier doping plays a crucial role. We also briefly examine 
\textit{non-local, hybrid} plasmons in an open system of a semi-infinite conductor, Coulomb-coupled to a 
2D layer in the presence of finite doping doping of the electrons in the layer. Our concluding 
remarks and a concise discussion are presented in Sec.~\ref{s4}. We also provide two appendices with detailed 
derivations of the DOS for silicene and $MoS_2$ - in Appendix \ref{apa}, and of the 
temperature-dependent chemical potential in Appendix \ref{apb}.

\section{Chemical potential}
\label{s2}

In this section, we discuss our analytical derivations for the electron chemical potential as a function of 
temperature. Being equivalent to the Fermi energy at $T=0$, the chemical potential normally decreases with   
increasing temperature. Its specific value depends on multiple material parameters such as energy band gaps,
Fermi velocities and the DOS of the of the electrons as well as the holes below the zero energy level. 
Thus, for a conventional 2D electron gas (2DEG) with no holes, the chemical potential could become 
negative at a certain temperature, which is not possible for a Dirac system with symmetry between
the electron and hole states. Here, we are going to provide closed-form trancendental equations for the 
finite-temperature $\mu(T)$ for a number of Dirac systems: graphene, buckled honeycomb lattices and transition 
metal dischacogenides.

\subsection{Buckled honeycomb lattices}

One of the most outstanding features of silicene and other buckled honeycomb lattices is the existence of two generally
double degenerate pairs of energy subbands and two inequivalent band gaps. These gaps There is a fixed intrinsic 
spin-orbit gap $2 \Delta_{SO}$ and a tunable sublattice-asymmetry gap $\Delta_z$  which is induced by and proportional 
to an applied perpendicular electric field $\mc{E}_z$. For small fields, $\Delta_z = \mc{E}_z \, d_{\bot}$, where 
$d_{\bot}$ is the out-of-plane displacement of a buckled lattice.

\par
\medskip
\par
The low-energy model Hamiltonian of a buckled honeycomb lattice has been found to be \,\cite{ezawa,SilMain} 

\begin{equation}
 \hat{\mbb{H}}_{\xi,\sigma} = \hbar v_F \left( 
 \xi k_x \hat{\tau}_x + k_y \hat{\tau}_y
 \right) \otimes \hat{\mbb{I}}_{2 \times 2} - \xi \Delta_{SO} \hat{\Sigma}_z \otimes \hat{\tau}_z +
\Delta_z \hat{\tau}_z \otimes \hat{\mbb{I}}_{2 \times 2} \, ,
\label{sil01}
\end{equation}
where the Fermi velocity $v_F = 0.5 \cdot 10^{6} \,m/s$ is half that for graphene, $\xi =  \pm 1$
is the $K/K'$ valley index, $\tau_{x,y,z}$ and $\Sigma_{x,y,z}$ are Pauli matrices in two different spaces,
attributed to pseudospin and real spin of the considered structure.

\par
\medskip
\par
Introducing spin index $\sigma = \pm 1$, we can rewrite Eq.~\eqref{sil01} in a block-diagonal matrix form

\begin{equation}
  \hat{\mbb{H}}_{\xi,\sigma} = \left( \begin{array}{cc}
                              - \xi \sigma \Delta_{SO} + \Delta_z & \hbar v_F (k_x - i k_y) \\
                                \hbar v_F (k_x + i k_y) & \xi \sigma \Delta_{SO} - \Delta_z  
                               \end{array} 
                               \right)
                               \, .
\end{equation}
The energy dispersions are

\begin{equation}
 \varepsilon_{\xi,\sigma}^{\gamma}(k) = \gamma \sqrt{
 \left(
 \xi \sigma \Delta_z - \Delta_{SO} 
 \right)^2 +
 \left(\hbar v_F k \right)^2 
 } \, ,
 \label{sildisp}
\end{equation}
where $\gamma = \pm 1$ determines the electron or hole state similar to graphene, with a finite or zero band gap.
These dispersions, given by Eq.~\eqref{sildisp}, represent two pairs of spin-dependent energy subbands in
a chosen valley with the two generally different band gaps $\vert \Delta_{SO} - \xi \sigma \Delta_z \vert $,
which will be later referred to as $\Delta_{<,>} = \vert \Delta_{SO} \mp \Delta_z \vert$. 
Clearly, both energy gaps depend on the perpendicular electrostatic field and the two subbands corresponding to the
$\xi \sigma = \pm 1$ indices switch their locations with increasing field strength. Small or zero $\mc{E}_z$ is related 
to a topological insulator (TI) states with $\Delta_z < \Delta_{SO}$. Once the two gaps become equal, we observe a 
metallic gapless state with $\Delta_{<}=0$ and a finite $\Delta_{>}$, defined as valley-spin polarized metal (VSPM).
For larger fields, $\Delta_z \backsim \mc{E}_z$ would always exceed the constant intrinsic spin-orbit gap $\Delta_{SO}$, 
which corresponds to the standard band insulator (BI) state.

\begin{figure}
  \centering
  \includegraphics[width=0.49\textwidth]{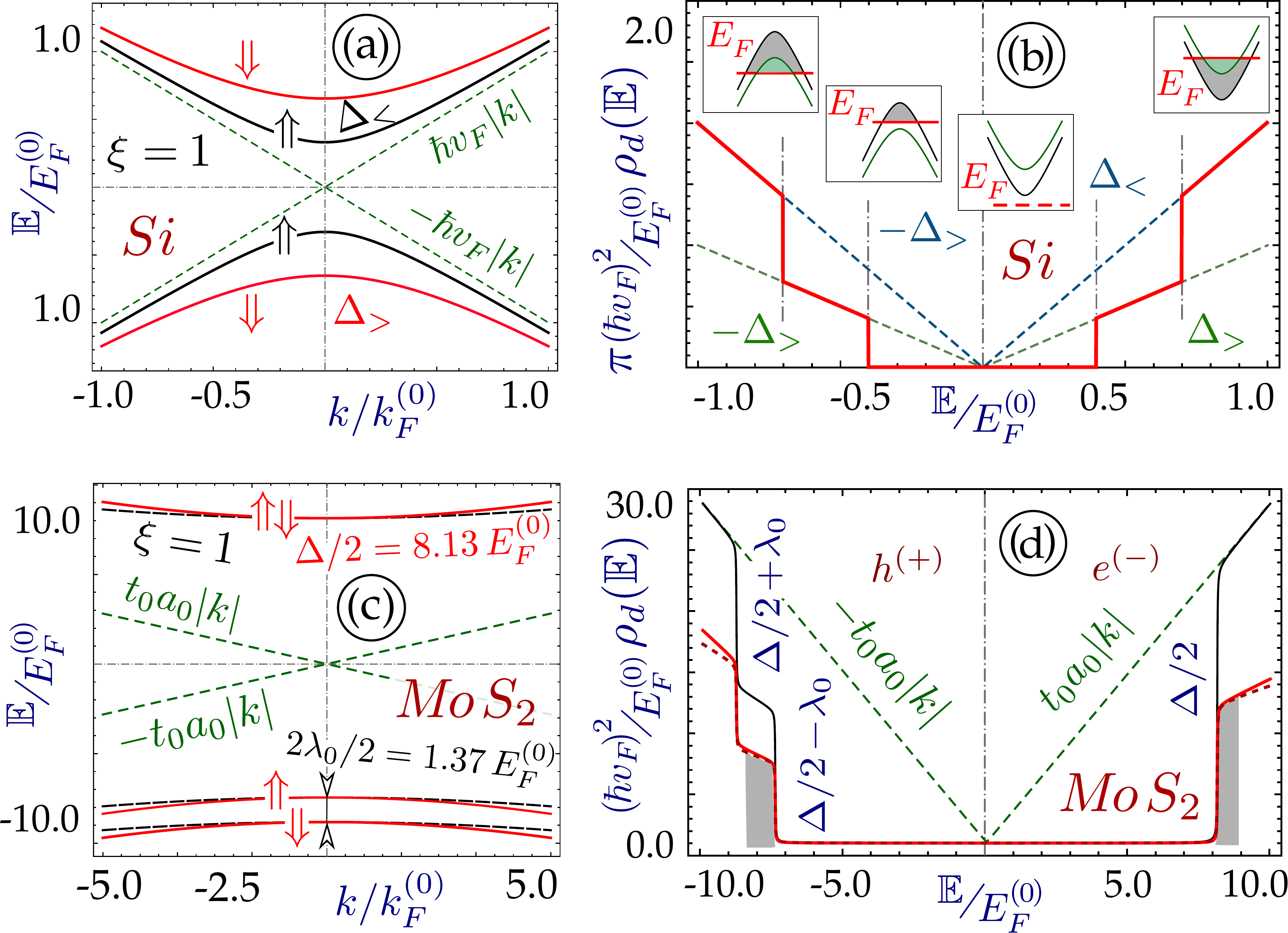}
  \caption{(Color online) Energy dispersions and density-of-states (DOS) for silicene (upper panels $(a)$ and 
  $(b)$) and molybdenum disulfide (plots $(c)$ and $(d)$). Left panels $(a)$ and $(c)$ represent low-energy
  dispersions in $K$ valley $(\tau = 1)$ for both materials. Linear dispersions, corresponding to zero 
  band  gap, are also provided for comparison. The range of the wave vector is extended to $ \pm 5 k_F^{(0)}$
  for $MoS_2$, in plot $(c)$, in order to display finite curvature of the dispersion curves which is markedly 
  suppressed due to a large band gap parameter $\Delta = 1.9 \, eV$. The DOS for TMDC's represented 
  in panel $(d)$, is calculated for the gapped graphene approximation, given by Eq.~\eqref{mo0}, and for a general
  model \eqref{ksq}, which displays substantial difference between the two results.  
  } 
  \label{FIG:1}
 \end{figure}

\par
\medskip
\par

The DOS which is in general defined by

\begin{equation}
\rho_d(\mbb{E})= \int \frac{d^2{\bf k}}{(2 \pi)^2} \sum\limits_{\gamma = 
\pm 1}\,\sum\limits_{\xi,\sigma= \pm 1} \, \delta\left[ \mbb{E} -
\varepsilon_{\xi,\sigma}^{\gamma}(k) \right] \, ,
\label{dos}
\end{equation}
is immediately obtained for silicene (see Appendix \ref{apa}) as

\begin{equation}
 \rho_d(\mbb{E}) = \frac{1}{\pi} \, \sum\limits_{\gamma = \pm 1}  \frac{\mbb{E}}{\hbar^2 v_F^2} 
\sum\limits_{i = <,>} \Theta \left[ 
\frac{\mbb{E}}{\gamma} - \Delta_i
 \right] \, , 
\end{equation}
in terms of the unit step function $\Theta(x)$. 
We note that for systems sharing the same Dirac cone characteristics with arbitrary energy gap, 
the DOS is linear analogous  to  graphene. Experimentally obtained linear V-shaped DOS was used 
to verify the Dirac cone dispersion for germanene. \cite{walh} However, $\rho_d (\mbb{E})$ has a 
finite value only above the energy gap since only for this energy range electronic states exist 
and tha  is how the band gap plays an important  role.

\par
\medskip
\par

Finite-temperature chemical potential for an electronic system is obtained using conservation 
of carrier density $n$  of electrons ($n^{(e)} $) and holes $ n^{(h)} $ concentrations at all 
temperatures, including $T=0$. In this regard, we have

\begin{equation}
 n = n^{(e)} + (-1) n^{(h)} = \int\limits_0^{\infty} d \mbb{E} \, \rho_d(\mbb{E}) f_{\gamma = 1}(\mbb{E},T) - 
 \int\limits_{-\infty}^{0} d \mbb{E} \, 
\rho_d( \mbb{E} )  \left\{ 1 - f_{\gamma = 1}(\mbb{E,T}) \right\} \, .
\label{EConsM}
\end{equation}

At zero temperature, the density $n$ is related to the Fermi energy $E_F$ in a straightforward way.
If only one subband is occupied, we obtain
\begin{equation}
 n = \frac{1}{2 \pi} \, \frac{E_F^2 - \Delta_<^2}{\hbar^2 v_F^2} \, .
 \label{n1M}
\end{equation}
Alternatively, if the doping density is such that both subbands are populate, then $E_F$ is obtained from

\begin{equation}
n = \frac{1}{\pi} \, \frac{1}{\hbar^2 v_F^2} \, \left[ E_F^2 - \frac{1}{2} \, \left( \Delta_<^2 + \Delta_>^2\right) \right] \, ,
\label{n2M}
\end{equation}

and the critical density required to start filling the upper subband is $n_c = 2 \Delta_{SO}\Delta_z / \pi \hbar^2 v_F^2$.

\par
\medskip
\par

We prove in Appendix \ref{apb} that  the finite-temperature chemical potentail could be obtained from the following
equation \

\begin{equation}
n \, \left( \frac{\hbar v_F}{k_B T} \right)^2 =  \sum\limits_{\gamma = \pm 1} \, \frac{\gamma}{\pi} \, \sum\limits_{i = <,>}
- \text{Li}_{\,2} \left\{ - \tet{exp} \left[\frac{ \gamma \mu(T) - \Delta_i}{k_B T} \right] \right\}  +
\frac{\Delta_i}{k_B T} \,  \ln \left\{ 1 + \tet{exp} \left[\frac{\gamma \mu(T) - \Delta_i}{k_B T}\right] 
\, \right\} \, ,
\label{musilM}
\end{equation}
where $\text{Li}_{\,2} (x)$ is a polylogarithm function. Connecting the doping density $n$ with the Fermi energy through either 
Eq.~\eqref{n1M} or Eq.~\eqref{n2M} depending on how many subbands are doped at zero temperature, we derive the 
chemical potential with a value equal to $E_F$ at $T = 0$, for all accessible  temperatures.
Although this equation is \textit{transcendental} and cannot be resolved algebraically, a quasi-analytic or 
one-step numerical solutions could be easily provided for any finite temperature without having to perform an integration.

\par
\medskip
\par
Clearly, the chemical potential for silicene depends on the energy band gaps $\Delta_i$, $i = < ,> $. Our approach, discussed 
in Appendix \ref{apb}, is valid for a variety of materials with linear energy dependence for the DOS, including $MoS_2$.  
Specifically, Eq.~\eqref{musilM} also describes $\mu(T)$ for gapped graphene with two degenerate subbands, or $\Delta_< = 
\Delta_> = \Delta_0$. For gapless pristine graphene $\Delta_0 = 0$ and $\pi n = \left[ E_F/ (\hbar v_F) \right]^2$, we have 

\begin{equation}
 \frac{1}{2 \left( k_B T \right)^2} \, E_F^2  
 = - \sum\limits_{\gamma = \pm 1} \gamma \,\text{Li}_{\,2} \left\{ - \tet{exp} \left[\frac{ \gamma \, \mu(T)}{k_B T}
 \right] \right\} 
 \, .
 \label{muG}
\end{equation}
If the temperature is low with $k_B T \ll E_F$, Eq.~\eqref{muG} is reduced to the expressions,  
derived in Refs. [\onlinecite{SDSS, SDSLi}] and [\onlinecite{myT}].

  \begin{figure}
  \centering
  \includegraphics[width=0.49\textwidth]{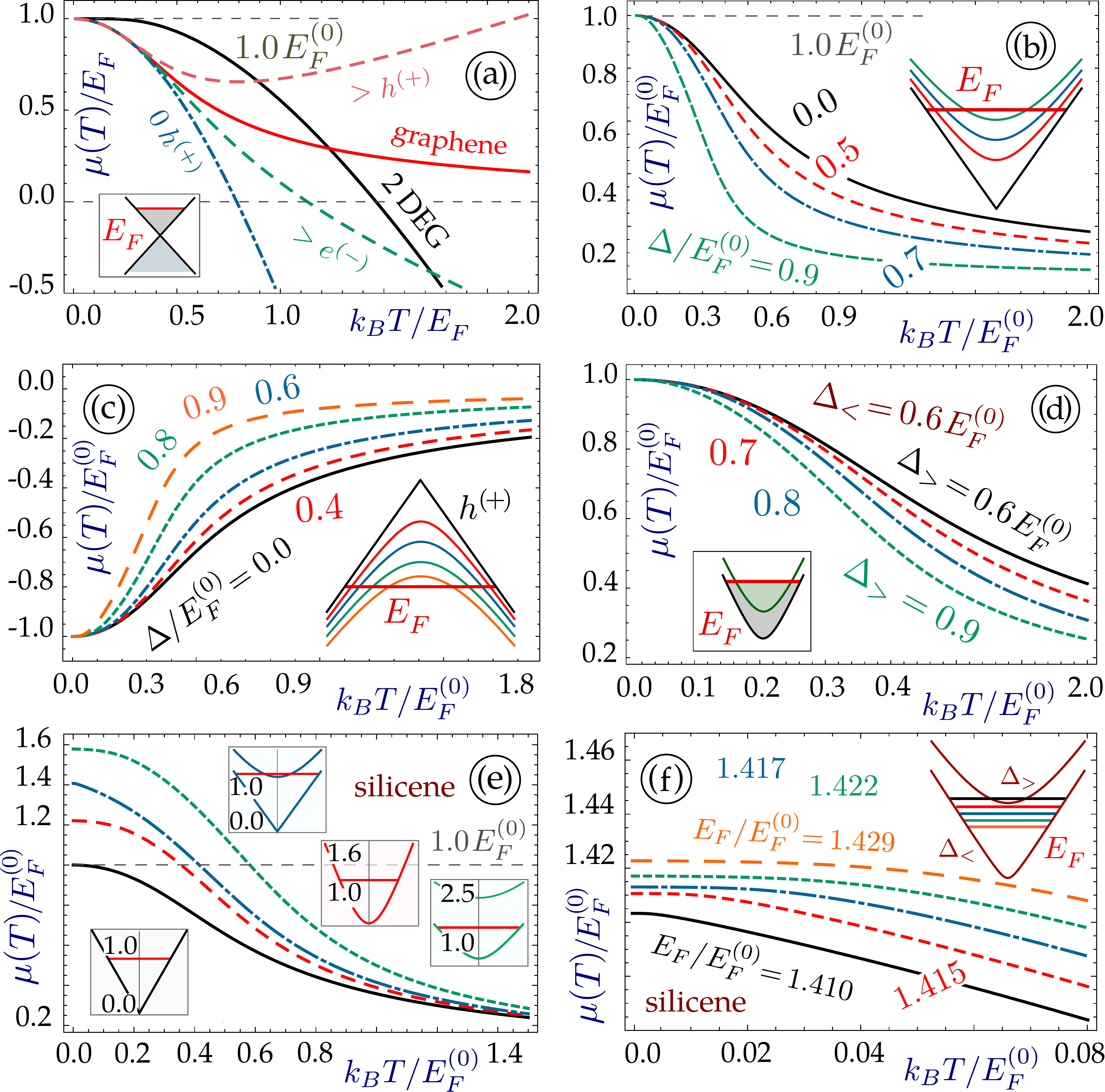}
  \caption{
  (Color online). Finite temperature chemical potential $\mu(T)$ for graphene and silicene. 
Panel $(a)$ shows $\mu(T)$ dependence for a number of representative cases - 2DEG with 
$\varepsilon \backsim k^2$, $\rho_d(\mbb{E}) = \text{const}$ and no holes (black solid line),
gapless graphene with $\rho_d(\mbb{E}) \backsim \mbb{E}$ (solid  red curve) and three others 
with broken electron-hole symmetry. Dash-dotted blue line describes graphene without holes 
$\rho_d^{(h)} (\mbb{E} < 0) = 0$, green dashed curve corresponds to graphene with reduced hole 
DOS $\rho_d^{(h)} (\mbb{E} < 0) = 1/2 \, \rho_d^{(e)} (\mbb{E} > 0)$, and the watermelon dashed line
shows the opposite situation, i.e.,   reduced electron DOS $\rho_d^{(e)} (\mbb{E} < 0) = 1/2 \, \rho_d^{(d)} 
(\mbb{E} > 0)$.
Plots $(b)$ and $(c)$ present $\mu(T)$ for gapped graphene for electron and hole doping, respectively, for  
chosen Fermi energy. Each curve is matched to a specific energy gap, as described schematically in the insets. 
Chemical potential for silicene with both subbands filled   $\Delta_< = 0.6\,E_F^{0}$ and different $\Delta_> $ 
is presented in panel $(d)$. Black  curve describes $\Delta_> = \Delta_< = 0.6\,E_F^{(0)}$ (gapped graphene), 
red and short-dashed line - $\Delta_< = 0.7\,E_F^{(0)}$, dotted blue line - $\Delta_< = 0.8\,E_F^{(0)}$ and dashed green
curve corresponds to  $\Delta_< = 0.9\,E_F^{(0)}$.  
Plot $(e)$ combines two degenerate-subband  situations - graphene  ($\Delta_{SO} = \Delta_z = 0.0$ and, $E_F = E_F^{(0)}$), 
black solid curve and gapped graphene with  $\Delta_{SO} = 0.7\, E_F^{(0)}$, $\Delta_z = 0.0$ and $E_F = 1.22 \, E_F^{(0)}$
(red dashed curve) with the two cases of silicene with split subbands  - $\Delta_{SO} = \Delta_z = 0.7\, E_F^{(0)}$ and 
$\Delta_{z} = 0.7\, E_F^{0}$, $E_F = 1.41 \, E_F^{(0)} \gtrapprox \Delta_>$, and , finally,  $\Delta_{SO} = 0.7\, E_F^{(0)}$,
$\Delta_z = 1.4 \, E_F^{(0)}$ and $E_F = 1.58\,E_F^{(0)}$. The subband schematics and the Fermi level for each case are given
in the insets.
Panel $(f)$ shows $\mu(T)$ at low temperatures satisfying $k_B T \leq 0.1\,E_F^{(0)}$ for a number of cases of doping
level located close to the upper subband $\Delta_> \approx E_F$. Here, $\Delta_{SO} = 0.7\,E_F^{(0)}$ for each curve, 
$\Delta_z = 0.70$, $0.75$, $0.80$ and $0.85\,E_F^{(0)}$, while the Fermi levels at $T=0$ are $1.410$, $1.415$, $1.417$, 
$1.422$ and $1.428\,E_F^{(0)}$. 
} 
  \label{FIG:2}
 \end{figure}

\par
\medskip
\par

All our considered materials could be effectively classified by the existing or broken symmetries of certain kinds, 
and, consequently, by the degeneracy of their energy subbands, which may be generally different for electrons 
and holes. In this respect, graphene represents the simplest case with a fourfold spin and valley degeneracy of 
$\pm \sqrt{(\hbar v_F k)^2 + \Delta_0^2}$ states. Silicene and germanene dispersions, yet showing complete electron/hole
symmetry, exhibit spin- and valley-dependent pairs of subbands, each being double degenerate. Finally, $MoS_2$ demonstrates
broken symmetry between its electrons and holes, and a finite energy separation between two non-equivalent holes subbands.

\par
\medskip
\par

Typical energy dispersions and DOS for silicene and molybdenum disulfide are presented in Fig.~\ref{FIG:1}.
For both materials, we consider the $K$ valley with $\xi = 1$ so that the upper electron and lower hole subbands correspond to
$\sigma = 1$ spin. Every time a new subband, or their degenerate manifold,   begins to be doped, we see an immediate increase.
or discontinuity, of the DOS, as schematically shown for silicene in the insets of Fig.~\ref{FIG:1} $(b)$. It is important 
to observe that for both silicene and gapped graphene, $\rho_d (\mbb{E})$ is directly proportional to energy $\mbb{E}$, i.e.,
the DOS for Dirac materials with finite and zero gap are exactly the same as long as we measure it above the 
enery band gap. With no electronic states inside the gap region, we have $\rho_d( \mbb{E} < 0) = 0$.

\par
\medskip
\par

In Fig.~\ref{FIG:2}, we display our results for the finite-temperature chemical potential for graphene and buckled 
honeycomb lattices. First, we show how dissimilar this temperature dependence might be for various electronic 
systems. In Fig.~\ref{FIG:2} $(a)$, this situation is described for a 2DEG with a parabolic energy band and constant
DOS, graphene with $\rho_s(\mbb{E}) \backsim \mbb{E}$, as well as two \textit{model structures} - graphene 
with no hole states, with doubly prevailing electrons $\rho_d(\mbb{E} > 0) = 2 \rho_d (\mbb({E} < 0)$ and doubly prevailing 
hole DOS. At low  temperatures, all the curves, except the one for the 2DEG, are nearly identical since the 
holes do net play any important role. The hole distribution function is complimentary to that for electrons, i.e., 
$1-f_e(\epsilon) \rightarrow 0$, and, therefore, inconsequential. 
However, when the temperature becomes comparable with the Fermi energy, the hole thermal excitations become crucial, 
causing an opposite effect compared with that for electrons. They mitigate the reduction of the chemical potential 
and eventually prevent $\mu(T)$  from  crossing the zero energy level. This is seen particularly well for a 
hole-dominating system $(> h^{(+)})$, for which the chemical potential starts to increase and ultimately exceeds the 
initial $E_F$ value.  We conclude that only total electron/hole symmetry, but not the energy gap, keeps the chemical 
potential positive for arbitrary high temperatures.

\par
\medskip
\par
For the remaining plots, we consider the behavior of $\mu(T)$   for graphene and silicene with different gaps. At $T=0$, we may 
keep the Fermi energy fixed so that the actual electron density $n$ differs,   in which case  the states with a larger 
gap receive much smaller amount. Alternatively,  we can dope the sample and  $\mu(T)$ shows much stronger reduction,
or we can fix the carrier density $n$  so that the Fermi energy 
increases in  the case with larger gap (see Eq.~\eqref{n1M}). The former case is shown in panels $(b)$, $(c)$ and $(d)$,
whereas the latter at $(e)$ and $(f)$. In general, the carrier density $n$  is accepted to be the most meaningful parameter, 
determining the Fermi energy for each specific system.

\par
\medskip
\par
Plots $(b)$ and $(c)$ are designed to demonstrate complete reflection symmetry of the chemical potential between electron 
and hole doping for all energy band gaps. This is a manifestation of $\gamma = \pm 1$ symmetry of these states persisting
at arbitrary temperatures. Silicene with both filled subbands exhibits qualitatively similar finite-temperature behavior as 
gapped graphene. More generally, only the actual gaps are relevant, but not the exact type of states (such as topological 
insulator or regular band insulator), or the way in which such a state has been achieved.

\par
\medskip
\par
We now pay attention to the following situation in silicene. At $T=0$, the Fermi level is chosen so that only the lower
subband is filled. In contrast, the upper one $\Delta_>$ is located so close  to the Fermi level $E_F$, that it starts 
getting populated at all, even very small temperatures. Thermally-induced doping, received by the $\Delta_>$-subband, results
in a much stronger reduction rate of $\mu(T)$, as shown in Fig.~\ref{FIG:2} $(e)$. Such situation suggests very special 
thermal properties of silicene with closely located energy subbands, and shares this behavior with graphene having an additional spin and
valley degeneracy. We investigate this phenomenon even further by adjusting the upper subband in the vicinity of the Fermi
level, as shown in Fig.~\ref{FIG:2} $(f)$. Now, each $\mu(T)$ curve demonstrates a significantly pronounced decrease whenever the 
upper subband filling becomes essential. These two different types of $\mu(T)$ behavior in silicene could be used to achieve 
its additional tunability by introducing the upper energy subband $\Delta_>$.

\subsection{Transition metal dichalcogenides}

The low-energy electronic states in monolayer molybdenum disulfide ($MoS_2$ , ML-MDS), a prototype transition metal 
dichalcogenide, could be effectively described by a \textit{two-band} model Hamiltonian \,\cite{xiao-prl,Scholz1,Habib}

\begin{equation}
\hat{\mbb{H}}_{d}^{\xi,\sigma} = \left( 
\frac{1}{2}\,\xi \sigma \, \lambda_0 + \frac{\hbar^2 k^2}{4 m_e} \alpha 
\right) \hat{\mbb{I}}_{2\times 2}
+
\left(
\frac{\Delta}{2} - \frac{1}{2}\,\xi \sigma \, \lambda_0 + \frac{\hbar^2 k^2}{4 m_e} \beta 
\right) \hat{\Sigma}_z + 
t_0 a_0\,\hat{\boldsymbol{\mathrm{\Sigma}}}_\xi \cdot {\bf k} \, ,
\label{mosham}
\end{equation}
whose important feature is a major  gap parameter $\Delta = 1.9\,eV$ which results in the actual 
band gap $\backsimeq 1.7 \, eV$,  large compared to that for silicene. The spin-orbit coupling 
parameter $\lambda_0=0.042\,\Delta$ represents a  smaller,
but essential correction, to the single-particle excitation spectrum and the band gap. The energy subbands are now spin- and valley-
dependent since the corresponding degeneracy is lifted. The electron hopping parameter $t_0 = 0.884\,\Delta$ and $a_0=1.843\,$\AA\ shape 
the Dirac cone term in the Hamiltonian Eq~.\eqref{mosham} as $t_0 a_0= 4.95\times 10^{-29}\,$J$\cdot m$, 
counting up to $\approx 0.47$  of $\hbar v_F$ value in graphene.

\par
\medskip
\par
Next, we include  the $\backsim k^2$ mass terms with $\alpha =2.21 = 5.140\,\beta$ in which $m_e$ is the
 free electron mass. Our considered values of Fermi momentum at zero temperature are determined 
by the experimentally allowed electron and hole doping densities $n = 
10^{14} \div 10^{16}\,m^{-2}$ as $k_F = \sqrt{\pi n} \backsimeq 10^{8} \div 10^{9}\,m^{-1}$.
\,\footnote{Expression $k_F = \sqrt{\pi n}$ holds true only for systems with a fourfold spin and 
valley degeneracy, such as graphene, 
i.e., $g = g_s g_v = 4$. The general equation for the Fermi momentum in a  2D material reads 
$ (2 \pi)^2 \, n = g \, \pi k_F$.}
Anisotropic trigonal warping term $t_1 a_0^2\,(\hat{\boldsymbol{\mathrm{\Sigma}}}_\xi
\cdot {\bf k})\,\hat{\sigma}_x (\hat{\boldsymbol{\mathrm{\Sigma}}}_\xi\cdot {\bf k})$ is clearly 
beyond out consideration since 
$\backsim t_1 = 0.1\,eV = 0.053\,\Delta$ term does not cause any effect on the electron dispersions.   
The energy dispersion relations, corresponding to the Hamiltonian 
Eq.~\eqref{mosham} $\varepsilon^{\xi,\sigma}_{\gamma}(k) = \mbb{\epsilon}_0^{\xi,
\sigma}(k) + \gamma \sqrt{ \left[ \Delta_0^{\xi,\sigma}(k) \right]^2 + (t_0 a_0 k)^2}$ , 
formally represent   \textit{gapped graphene} with 
a $k-$dependent gap term  $\Delta_{0}^{\xi, \sigma}(k) = \hbar^2 k^2 \beta / (4 m_e)
 + \Delta / 2 - \xi \sigma \, \lambda_0 / 2 $, and a band
shift $\mbb{\epsilon}_0^{\xi,\sigma}(k) = \hbar^2 k^2 \alpha /(4 m_e) + \xi \sigma \, 
\lambda_0 / 2$. \,\cite{Scholz1, ourjap2017}

\par
\medskip
\par

Neglecting only the ${\cal O}(k^4)$ terms leads us to 

\begin{equation}
\varepsilon^{\xi,\sigma}_{\gamma}(k)\backsimeq \frac{1}{2}\,\xi \sigma \lambda_0 + \frac{\alpha \hbar^2}{4 m_e}\,k^2 + 
\frac{\gamma}{2}
\left\{
\left[
(2 t_0 a_0)^2+\left( \Delta - \xi \sigma \, \lambda_0  \right)  \beta\hbar^2 / m_e
\right] k^2 +
\left( \Delta - \xi \sigma \, \lambda_0 \right)^2
\right\}^{1/2} \, .
\label{ksq}
\end{equation}
This is the principal model which we will use to describe the energy dispersions of $MoS_2$ in our work. 
The $\backsimeq k^4$ terms, trigonal warping and anisotropy are considered to be non-essential, even though
 as we will later see, cause certain discrepancies in the DOS. 
\footnote{Due to a large  gap parameter $\Delta = 1.9\,eV$, the electronic states corresponding to large 
wave vectors are often involved and at $k \approx 5.0\,k_F^{(0)}$ correction 
 $\backsimeq \Delta \, \beta \,\, k^4$ is no longer negligible.}
We also show (see Appendix \ref{apa}) that the curvature of the energy subbands in TMDC's is so small that
 even the highest  possible doping density $10^{17}\,m^{-2}$ results in the Fermi energies 
$\backsim \lambda_0$. Thus, at zero or low temperatures, we do not need to consider any high-energy 
corrections to Eq.~\eqref{ksq}  On the other hand, inclusion of higher 
order terms into the dispersions, would enormously complicate the DOS calculation.

\par
\medskip
\par
At high temperatures,  the  electronic states far from the Dirac point would also receive 
substantial temperature-induced doping density due to the so-called Fermi tail. \,\cite{myT} In
that case, our model Hamiltonian in Eq.~\eqref{mosham} and especially, simplified dispersion 
relation \eqref{ksq} would no longer provide a satisfactory   approximation. Consequently,
our primary focus is on \textit{small but finite temperatures} for which the initial doping density and $E_{F}$ still play an 
important role, beyond the ${\cal O} \left(T^2/T_F^2 \right)$ approximation discussed in Ref.~[\onlinecite{SDSLi}].

\par
\medskip
\par
Here we note that for $MoS_2$, similar to the buckled honeycomb lattices, spin and valley indices always 
appear together as a product, so that taking into account a $\times 2$ degeneracy, a single composite index
$\nu = \xi \sigma$ could be effectively introduced. We are going to use only the index $\nu$ for the rest of our 
consideration.    A valley- and spin-resolved gapped graphene approximation of dispersions \eqref{ksq} arises 
once we also neglect the  mass terms in Eq.~\eqref{ksq} \,\cite{ggmosXiao}

\begin{equation}
\varepsilon^\nu_{\gamma}(k)\backsimeq \nu\lambda_0/2+\gamma\sqrt{(t_0 a_0)^2 k^2+(\Delta-\nu\lambda_0)^2/4} \, .
\label{mo0}
\end{equation}
This approximation has a few important advantages  including simplicity and its formal resemblance 
with gapped graphene so that all the crucial quantities such as the DOS, wave function,  polarizability 
and many others are already known. Furthermore, it gives a quite a suitable description of the energy 
band structure of $MoS_2$, taking into account a large gap parameter and $\nu$-dependent splitting of the two hole subbands.  
Nevertheless, the mass terms must be taken into account for a proper evaluation of the DOS and most of the 
temperature-dependent properties of TMDC's. As we demonstrated in Appendix \ref{apa}, even in the simplest possible parabolic
subbands approximation, valid only for small wave vectors $k \rightarrow 0$, the mass terms make a contribution,
comparable with the Dirac cone and band gap parts of Hamiltonian \eqref{mosham}. Very small curvature of the energy subbands, 
mentioned above results in a tremendous $\rho_d \backsim m_{eff}$, so that even $\backsimeq \alpha/(4 m_e)$ correction, hardly
noticeable on the electron band structure,   becomes significant. It basically discards the DOS results obtained 
from the gapped graphene model, even at zero or small wave vectors. Importance of the mass and even higher order terms for the
plasmon calculation was discussed in Ref.~[\onlinecite{Scholz1}].

\par
\medskip
\par
Taking into account all the required terms in \eqref{mosham},   rigorous numerical calculations give   
accurate results for the electron DOS for $MoS_2$. In Fig.~\ref{FIG:1} $(d)$, we present all three possible
outcomes.  Based on the gapped graphene model, the DOS is nearly twice as large as its numerical values
. \,\cite{MoS238, ourjap2017} In contrast, our $\backsim k^2$ model \eqref{ksq} demonstrates quite a good match,
 especially in the low-energy range.   We also note that the numerically obtained dependence is clearly linear 
for a wide range of energies, much exceeding our considered diapason.

\par
\medskip
\par
In summary, we consider a piecewise linear approximation 
$\rho_d (\mbb{E})$ relatively close $(\delta \mbb{E} \backsimeq \lambda_0 )$ for each of
the three non-degenerate subbands $\rho_d(\mbb{E}) = c_0^{(i)} + c_1^{(i)} \, \mbb{E}$ and $\rho_d(\mbb{E}) = 
0 $ in the gap region $-\Delta/2 + \lambda_0 < \mbb{E} < \Delta/2$. The expansion coefficients 
are obtained as $c_0^{(1)} =  2.837 \, E_F^{(0)}/(\hbar v_F)^2 = 0.043 \,t_0^{-1} a_0^{-2}$, $c_1^{(1)} = - 
1.397 \, (\hbar v_F)^{-2} = -0.308 \,(t_0 a_0)^{-2}$ 
for $\gamma = -1$, $\nu = -1$ and $\leq -\Delta/2 - \lambda_0 \geq \mbb{E} \leq  -\Delta/2 - \lambda_0$.
Finally, when $\gamma = -1$, but $\nu = 1$ and $\mbb{E} \leq -\Delta/2 + \lambda_0$,
the hole DOS coefficients are $c_0^{(2)} = 1.132 \, E_F^{(0)}/(\hbar v_F)^2 = 0.0174 \,t_0^{-1} a_0^{-2}$
$c_1^{(2)} = -0.767 \, (\hbar v_F)^{-2} = -0.169 \,(t_0 a_0)^{-2}$. For electrons with $\gamma=+1$ at 
$\mbb{E}= \Delta/2 + \delta \epsilon$, the two quasi-degenerate subbands lead to the DOS 
equal to $c_0^{(3)} = 5.110 \, E_F^{(0)}/(\hbar v_F)^2 = 0.078 \, t_0^{-1} a_0^{-2}$ and $c_1^{(3)} = 
0.815 \, (\hbar v_F)^{-2} = 0.179 \,(t_0 a_0)^{-2}$.

\par
\medskip
\par
We adopt these values for $\rho_d(\mbb{E})$, which arise from the numerical calculations in order 
to achieve the highest possible precision and credibility for our finite-temperature derivations. However, 
 our effective model, presented in Appendix \ref{apa}, gives the DOS results which show good agreement with 
these numerical values and could be used for decisive estimates for various collective calculations for $MoS_2$.

\begin{figure}
  \centering
  \includegraphics[width=0.49\textwidth]{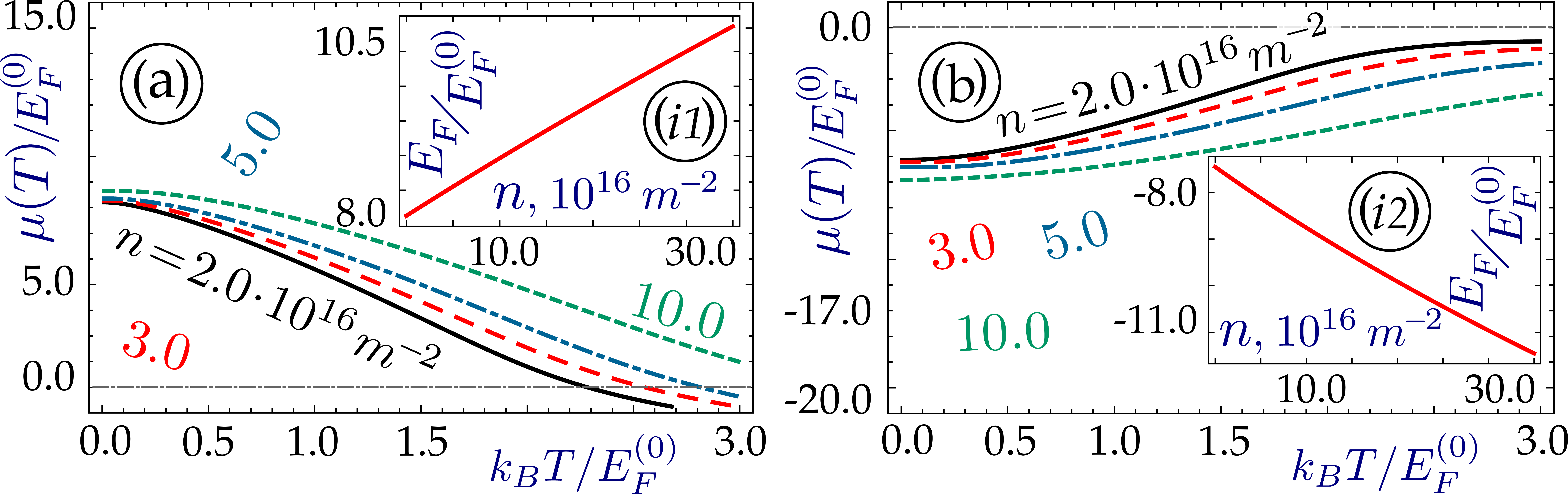}
  \caption{
  (Color online). Temperature-dependent chemical potential for MoS$_2$. Panel $(a)$ presents the $\mu(T)$ 
	dependence 	for various   initial \textit{electron} doping densities - $n=2.0 \cdot 10^{16}\,m^{-2}$ (black 
	solid line), $n=3.0 \cdot 10^{16}\,m^{-2}$ (red and dashed), $n=5.0 \cdot 10^{16}\,m^{-2}$ (dotted blue curve) 
	and $n=1.0 \cdot 10^{17}\,m^{-2}$ (green short-dashed line). Plot $(b)$ shows  the corresponsing dependence 
	for   \textit{hole doping} with the same densities. Insets $(i1)$ and $(i2)$ illustrates how the Fermi energy 
  depends on the electron and hole doping densities at zero temperature. 
  } 
  \label{FIG:3}
 \end{figure}

\par
\medskip
\par
Once the DOS is known, we are in a  position to calculate the Fermi energy for a given doping density $n_{(0)}^{e}$.  
The new point here is that $\rho_d (\mbb{E})$ is not directly proportional to the energy so that $E_F$ is determined by

\begin{equation}
 n_{(0)}^{e} = \left(
 E_F - \frac{\Delta}{2}
\right) \,
\left[ c_0^{(3)} +
\frac{c_1^{(3)}}{2} \, \left(
\frac{\Delta}{2} + E_F
\right)
\right] \, , 
\label{neMo}
\end{equation}
or 

\begin{equation}
 E_F^{(e)} = \frac{1}{2 \, c_1^{(3)}} \, \left\{-2 \,c_0^{(3)} + \left[ \left( 2 \, c_0^{(3)} + c_1^{(3)} \, \Delta \right)^2 + 8 n_{(0)}^{e} 
 c_1^{(3)} \right]^{1/2} \right\} \, .
\end{equation}
For hole doping, the Fermi energy differs from the previously considered electron doping case, i.e., 

\begin{equation}
 E_F^{(h)} =  \frac{1}{c_1^{(2)}} \, \left\{ 
 - c_0^{(2)} + \left\{ 
 -2 c_1^{(2)} \, n_{(0)}^{h} + \left[
 c_0^{(2)} - c_1^{(2)} \,
 \left(
 \frac{\Delta}{2} - \lambda_0
 \right)
 \right]^2 
 \right\}^{1/2} \,
 \right\}
\end{equation}
Numerically, our results for the  Fermi energy for electrons and holes are presented in insets 
$(i1)$ and $(i2)$  of Fig.~\ref{FIG:3}. Here, both linear and quadratic terms in the doping density 
equations are present (see Eq.~\eqref{neMo}), and, most importantly, \textit{there is no symmetry}
between the electron and hole states. Unlike graphene, the linear dependence here dominates for both 
cases due to the large energy band gap. Each curve starts from the corresponding band 
gap - $\Delta/2 = 8.13\,E_F^{(0)}$ for electrons and $-\Delta/2 + \lambda_0 = - 7.45\,E_F^{0}$. 
The corresponding well-known result for gapped graphene $ E_F^2 - \Delta_0^2 = 
2 \pi \, n \, \left(\hbar v_F \right)^2 $ are verified by putting $c_1^{(i)}
\rightarrow 0$ and $\lambda_0 \rightarrow 0$.

\par
\medskip
\par
The finite-temperature chemical potential for $MoS_2$ is obtained in a similar way, as we have 
done for the buckled honeycomb lattices, except that we need to evaluate \textit{four} different 
terms related to the two separate hole subbands (see Eq.~\eqref{Ihmo}).  The corresponding 
numerical results are described in Fig.~\ref{FIG:3}. As discussed above, the most special 
property of TMDC's is the broken electron/hole symmetry. Consequently, the chemical potential for 
the electron doping becomes negative at $T \approx 2.5\,E_F$, while $\mu(T)$ for hole  
doping does not ever change its sign. Broken electron/hole  symmetry leads to two substantially 
different types of temperature dependence of the chemical potential. Thus,   MoS$_2$ 
represents a special material with unique symmetry properties and chemical potential dependence, 
so far encountered only in certain types of semiconductors but not in Dirac materials.

\section{Plasmon Excitations at Finite temperature}
\label{s3}

\begin{figure}
  \centering
  \includegraphics[width=0.49\textwidth]{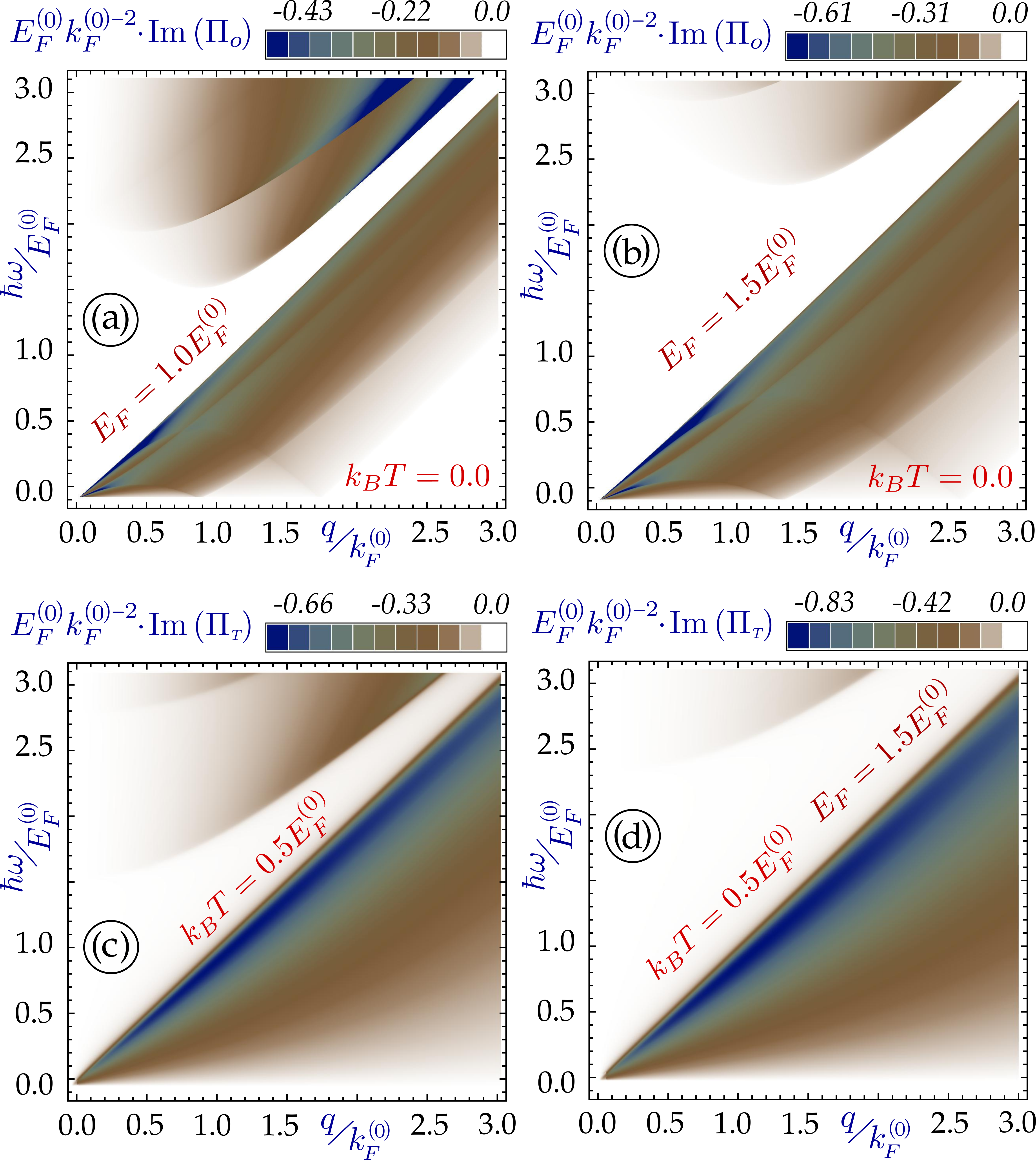}
  \caption{(Color online) Signle-particle excitation regions or particle-hole modes, outlined by non-zero 
  $\text{Im}\,\Pi_{T}(q,\omega)$ at an arbitrary temperature for silicene with $\Delta_{SO} = 0.7\,E_F^{(0)}$ 
  and $\Delta_z = 0.2\,E_F^{(0)}$. The upper panels $(a)$ and $(b)$ describe the situation for zero temperature, 
  while the lower plots $(c)$ and $(d)$ were obtained for $T=0.5\,E_F^{(0)}$. Left panels $(a)$ and $(c)$ are for  a 
  system with  $E_F = 1.0\,E_F^{(0)}$ whereas   the right ones - with $E_F = 1.5\,E_F^{(0)}$. Alternatively, the regions of
  $\text{Im}\,\Pi_{T}(q,\omega) = 0$ specify plasmon excitations with no Landau damping. 
  } 
  \label{FIG:4}
 \end{figure}

\begin{figure}
  \centering
  \includegraphics[width=0.49\textwidth]{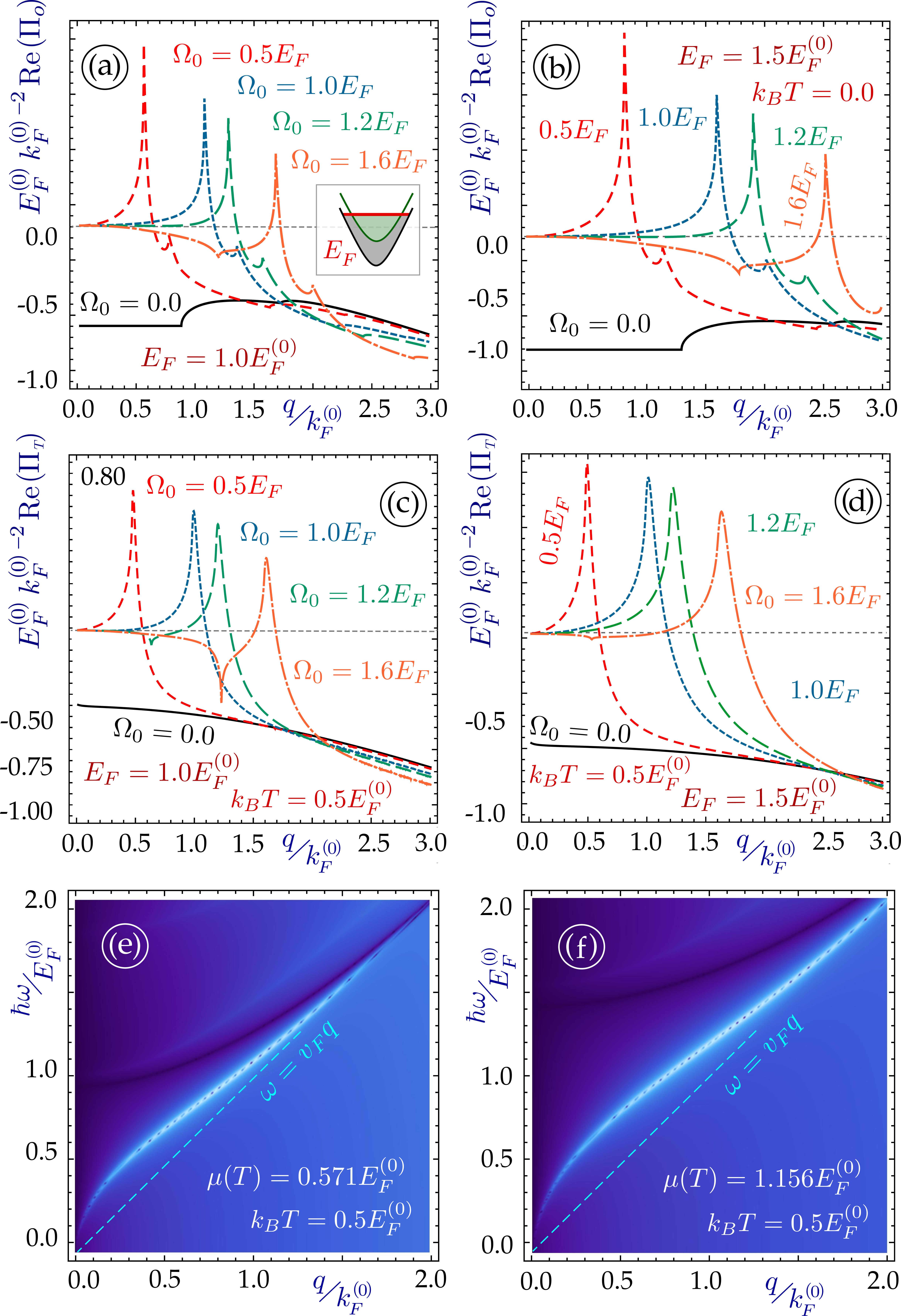}
  \caption{(Color online) Dynamic polarization function and plasmon excitations for silicene with 
	$\Delta_{SO} = 0.7\,E_F^{(0)}$    and $\Delta_z = 0.2\,E_F^{(0)}$. 
  Panels $(a)$ - $(d)$ present constant frequency cuts of $\text{Re}\,\Pi_{T}(q,\omega)$
  as a function of $q$ for $\Omega_0 =$ $0.0$, $0.5$, $1.0$, $1.2$ and $1.6 \, E_F^{(0)}/\hbar$, described by
  solid black, dashed red, short-dashed blue, long-dashed green and dash-dotted orange lines, respectively, 
  for each plot. Panels $(a)$ and $(b)$ show the zero temperature case, $(c)$ and $(d)$ - $T = 0.5\,E_F^{(0)}$. 
  Initial ($T=0$) doping is chosen to yield $E_F = 1.0\,E_FT^{(0)}$ for plots $(a)$ and $(c)$ while panels $(b)$ and $(d)$
  present the case of $E_F = 1.5\,E_FT^{(0)}$.
  Plasmon dispersions at $T = 0.5\,E_F^{(0)}/k_B$ for $E_F = 1.0 E_F^{(0)}$ is represented in plont $(e)$, 
  and for $E_F = 1.0 E_F^{(0)}$ - in panel $(f)$.  
  } 
  \label{FIG:5}
 \end{figure}

As one of the most relevant applications of our finite-temperature chemical potential formalism, 
we briefly consider plasmons for an extrinsic, substantially doped at $T=0$, free-standing gapped Dirac cone material.
The plasmon dispersion relation is calculated from the zeros of the dielectric function $\epsilon(q, \omega)$, 
expressed in the random phase approximation (RPA) as

\begin{equation}
\epsilon(q, \omega) = 1 - v(q) \, \Pi_{T}(q, \omega \, \vert \, \mu(T), T, \Delta_i) = 0 \, ,
\label{eps}
\end{equation}
where $v(q) = 2 \pi e^2 / (\epsilon_s q)$ is the Fourier-transformed two-dimensional Coulomb potential, 
and $\epsilon_s = 4 \pi \epsilon_0 \epsilon_r$ ith  $\epsilon_b$ is the background dielectric constant  
in which the 2D material is embedded.
At finite temperature, the dynamical polarization function $\Pi_{T}$ is given as an integral transformation \cite{malda}
of its $T=0$ counterpart $\Pi_{0}$, i.e., 

\begin{equation}
 \Pi_{T}(q, \omega \, \vert \, \mu(T), T, \Delta_i) = \frac{1}{2 k_B T} \, \int\limits_{0}^{\infty}  d \xi \,
 \frac{\Pi_{0} (q,\omega \, \vert \, \xi, \Delta_{i})}{1 + \cosh \left\{ [ \mu(T,E_F) - \xi] / (k_B T) \right\} } 
\label{Pi0T}
 \end{equation}
The evaluation of the zero-temperature polarizability is quite similar for buckled honeycomb lattices and $MoS_2$,
since in both cases their low-energy bandstructure is represented by two generally inequivalent double-degenerate
pairs of subbands which depend on the composite index $\nu = \sigma \, \xi$. For any such pair,   $\Pi_{0}^{(\nu)}$ is
obtained in the one-loop approximation ($g_0 = 2$) as

\begin{equation}
  \Pi_{0}^{(\nu)} (q,\omega \, \vert \, E_F, \Delta_{\nu}) = \frac{g_0}{8 \pi^2} \int d^2 k \sum\limits_{\gamma,\gamma' = \pm 1} \left( 
  1 + \gamma \gamma' \,
  \frac{{\bf k \cdot ({\bf k} + {\bf q})} + \Delta_\nu^2 }{ \varepsilon^{\nu} (k) \,\, \varepsilon^{\nu}(\vert { \bf k} + {\bf q} \vert) } \right)
  \frac{f[ \mbb{E} - \varepsilon^{\nu}_{\gamma} (k)] - f[\mbb{E} - \varepsilon^{\nu}_{\gamma'}(\vert { \bf k} + {\bf q} \vert)]}
  { \varepsilon^{\nu}_{\gamma} (k) - \varepsilon^{\nu}_{\gamma'} (\vert { \bf k} + {\bf q} \vert)} \, ,
  \label{pi00}
\end{equation}
where $f[\mbb{E} - \gamma \, \mbb{E}_{\nu}(k)]$ is the Fermi-Dirac distribution function, showing electron and hole 
occupation numbers for chosen energy $\mbb{E}$. At $T=0$, it is a Heaviside unit step function 
$\Theta \left[ \mbb{E} - \varepsilon^{\nu}_{\gamma}
(k) \right]$. We note that the extra 1/2   comes from the form factor $1/2 \, [ 1 + \gamma \gamma' ... $ 
 which also depends on the absolute value of each energy 
$\varepsilon^{\nu} (k) = 1 / \gamma \,\, \varepsilon^{\nu}_{\gamma} (k)$.  We also note that since any valley 
or spin transitions are inadmissible and only one summation over the index $\nu$ is incuded, compared to the     
electron/hole indices $\gamma$ and $\gamma'$.
Thus, for both types of materials, the dynamical polarization function is obtained as a sum of terms obtained from 
Eq.\  \eqref{pi00} over $\nu$  \,\cite{SilMain} 

\begin{equation}
 \Pi_0 (q,\omega \, \vert \, E_F, \Delta_{i}) = \sum\limits_{\nu} \, \Pi_{0}^{(\nu)} (q,\omega \, \vert \, E_F, \Delta_{\nu}) \, .
\end{equation}
Our results for the polarization functions and plasmon excitations are presented in Figs.~\ref{FIG:4} and \ref{FIG:5}. 

\par
\medskip
\par

First, we need to address the imaginary part of the polarizability  since it specifies the regions and intensity
of the plasmon damping. We see that at a finite temperature the plasmon dissipation generally increases and the 
damping-free regions are nearly absent. Once the temperature becomes   high, the imaginary part of the polarization
function is reduced as $1/T$, \,\cite{SDSLi,myT} so there is no uniform temperature dependence of the plasmon damping. 
Doping and proportional increase of the energy band gaps, in contrast, increase the regions free of single-particle 
excitation spectrum (see Fig.~\ref{FIG:4} $(b)$), so the effect of a finite temperature and $E_F$ on the plasmon dissipation
coul be opposite.
The real part of $\Pi_0 (q,\omega \, \vert \, E_F, \Delta_{i})$, shown in Fig.~\ref{FIG:5} $(a)$-$(d)$, designates the 
location and slope of the corresponding plasmon branches which are presented in panels $(e)$ and $(f)$. Here,
the temperature and the doping produce a similar effects, shifting the location of the peaks to the right.
The real part of the polarization function must be positive in order to enable a real solution of Eq.~\eqref{eps} 
Plasmon branches are located at higher energies for a given wave vector due to either an initial increase of the Fermi 
energy or thermally-induced doping. 

\subsection{Non-local plasmons in an open system}

Concluding our investigation of extrinsic 2D materials, we now turn to the     
plasmon excitations in so-called 2D open systems (2DMOS).
Such a nanoscale hybrid arrangement being a part of graphene-based nanoscale devices, \cite{ndbook1, ndbook2} consists
of a 2D layer (it could be 2DEG, graphene, a buckled honeycomb layer, or $MoS_2$), which is Coulomb coupled
(not chemically bound) with a semi-infinite conductor and its surface plasmon. \,\cite{surfplas, ourPRB15}
While a plasmon excitation in a closed system is determined by two-particle Green’s functions, 
in 2DMOS it more involved, depending on the  Coulomb interaction with the environment. \,\cite{open21, open32}
The important feature of such a system is the screened Coulomb coupling between the electrons in graphene and 
the conducting substrate.\cite{Gbook} Such a screened potential could be obtained using the nonlocal frequency-dependent 
inverse dielectric function \,\cite{hbook, nho1, nho2, nho3}  Consequently, two \cite{ourPRB15} or 
more \cite{ourSREP} linear plasmon branched have been obtained, confirming some previous experimental  
claims. \,\cite{Kramberger, poli1}   For finite temperature, their coupling to an external reservoir is
 reflected in existence of extra  plasmon dissipation channels. \cite{myT}

\par
\medskip
\par
The plasmon branches in such system are obtained as zeros of a so-called structure factor 
$S (q,\omega \, \vert \, \mu(T), \Delta_i )$, playing the role of the dielectric function 
$\epsilon (q, \omega)$ in an isolated layer. It is obtained as \,\cite{ourPRB15, myT}

\begin{equation}
 S (q,\omega \, \vert \, \mu(T), \Delta_i) = 1 - v(q) \,
 \Pi^{(0)}(q,\omega;\mu) \left\{
 \frac{1 - \epsilon_B(\omega)}{1 + \epsilon_B(\omega)} \, \tet{Exp}(-2 a \, q) + 1
 \right\} \, ,
\end{equation}
where $a$ is the distance between the 2D layer and the surface and the bulk dielectric function
is given in the local limit as $\epsilon_B = 1- \Omega_p^2/\omega^2$. The bulk-plasma frequency,
defined as $\Omega_p^2 = (n_m e^2)/(\epsilon_o \epsilon_S m^\ast)$, depends on on the electron concentration
$n_m$, its effective mass $m^\ast$ and the substrate dielectric constant $\epsilon_S$. This approximation 
stays valid for a large range of wave vectors $q \ll 2 \times 10^{9} m^{-1}$ since the Fermi 
wavelength in metals is comparable with the inverse lattice constant. As a result the frequency of the 
upper plasmon branch in our system, equal to $\Omega_p/\sqrt{2}$ at $q_\parallel \rightarrow 0$, might range from 
ultra violet down to infrared or even terahertz, depending on the substate material and must stay commensurate
with the energy band gap in the 2D layer. 

  \begin{figure}
  \centering
  \includegraphics[width=0.49\textwidth]{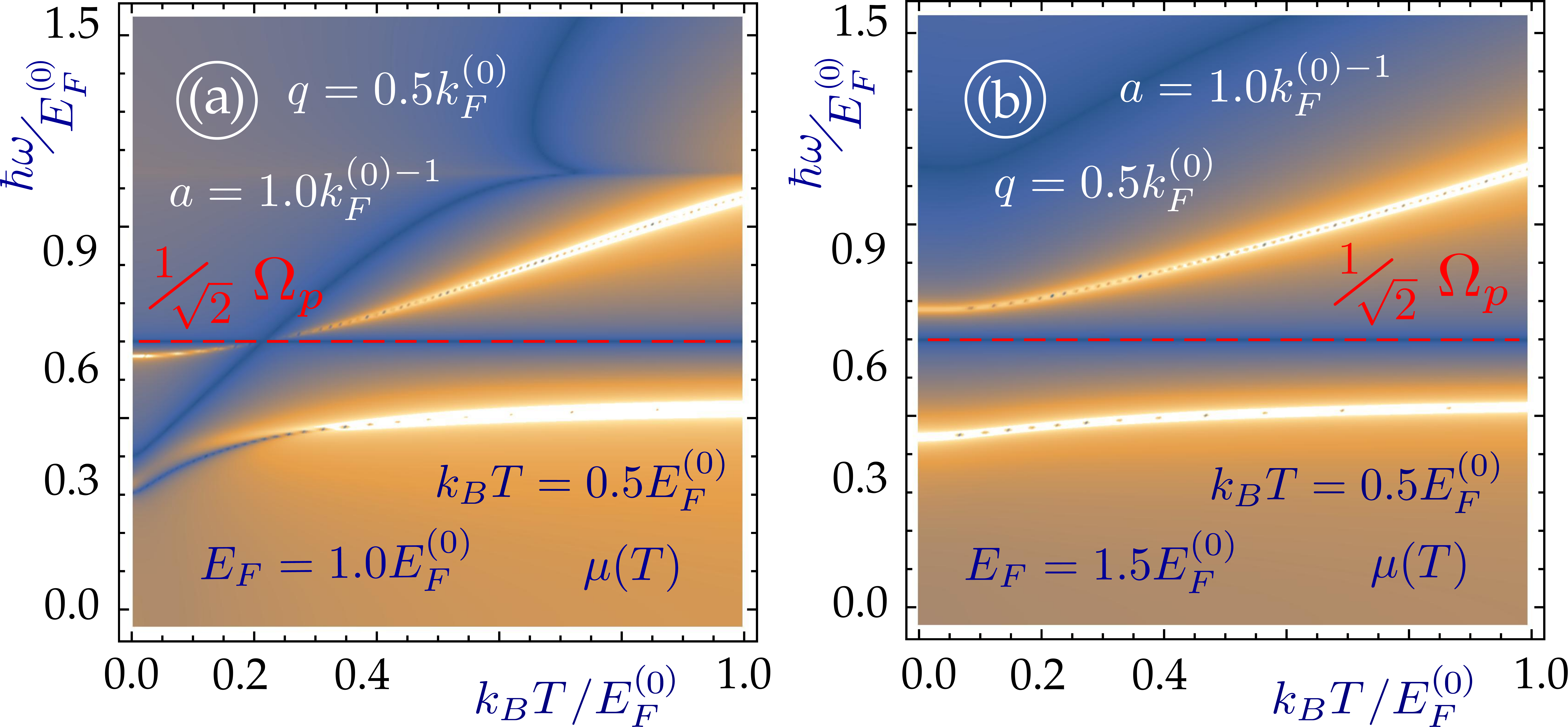}
  \caption{(Color online) Temperature-dependent non-local plasmon excitations for a silicene-based
  hybrid system with $\Delta_{SO} = 0.7\,E_F^{(0)}$ and $\Delta_z = 0.2\,E_F^{(0)}$. Density plots 
  of the real part of $\mathbb{S}_c(T, \omega+0^{+})$ for a constant wave vector $q = 0.3 k_F^{(0)}$ are
  presented, so that the sought low-damped plasmons correspond to their peaks. Panel $(a)$ corresponds to an
  intrinsic system with no zero-temperature doping, while plot $(b)$ is related to $E_F = 2.0\,E_F^{(0)}$.
  } 
  \label{FIG:6}
 \end{figure}

\par
\medskip
\par 

Previously, we reported that \,\cite{ourjap2017} in case of spin- and valey-dependent single-particle excitations
in a 2D layer (which is true for buckled honeyomb lattices and $MoS_2$ ), such a hybrid structure
could be effectively used to directly measure the dielectric properties or spin-orbit coupling parameters 
of such a layered material because the location of each plasmon branch, its damping rate and the signatures of the 
particle-hole modes are independently determined by the material parameters of the 2D layer. 
Each of these properties is unique (for example, the two plasmon branches in silicene-based TDMOS depends on 
the energy band gaps as $\Delta_i^{1/2}$ and $\Delta_i^{1/4}$, while the outermost PHM  boundaries are 
determined solely by the lower gap $\Delta_<$), so that an additional linear plasmon branch provides us 
with the required earlier unknown piece of information about the specific material. This is not possible 
in the case of a single plasmon branch in an isolated layer.

\par
\medskip
\par
In the present work, we additionally introduce finite doping and temperature into this consideration. Our numerical
results for the non-local plasmons for a extrinsic systems are presented in Fig.~\ref{FIG:6} We see that the location 
and damping of both branches substantially changes in the presence of  an initial Fermi energy. At low temperature, this
increase is especially apparent, similar to plasmons in an isolated layer. For a sample with stronger doping, 
$E_F = 1.5 E_F^{(0)}$ at zero temperature, the upper plasmon branch is always located above the surface plasmon level
$\Omega_p/\sqrt{2}$ and is never damped.

\section{Concluding remarks}
\label{s4}

We have carried out an extensive investigation of extrinsic, or doped, Dirac gapped  materials
at arbitrary finite temperatures and obtained a set of  algebraic analytic equations determining the chemical 
potential. Our considered systems include graphene, with or without an energy band gap, buckled honeycomb 
lattices with spin- and valley-dependent energy subbands and reduced degeneracy, as well as the  transition metal 
dichalcogenides having broken symmetry between the electron and hole states. Our results could also be used 
to predict the finite-temperature chemical potential for parabolic or quasi-parabolic eigenstates in 
semiconductors with light or heavy holes. \,\cite{semi1, semi2} In general, our model is limited
only by the linear dependence of the DOS which stays valid over a wide energy range for all the 
above mentioned materials.

\par
\medskip
\par

We have demonstrated that the chemical potential depends substantially on the energy band gap(s) of the considered 
system since the DOS depends on the curvature of each subband. Specifically, we investigated structures
with two non-degenerate, separated spin- and valley- resolved energy subbands in both valence and conduction bands, 
such as silicene. The upper subband would receive thermally-induced doping even if it is undoped at zero temperature. This is
always reflected in a higher reduction rate of the chemical potential whenever the second subband doping starts playing 
a role. Consequently, one can tune the $\mu(T)$ dependence around any required temperature by bringing initial doping close
to the higher-energy subband. The number of such separated subbands contributing to the DOS could be arbitrarily large for 
an electron in a quantum well or quasi-one-dimensional nanoribbons, \,\cite{ezawasnr} and all these cases could be effectively
treated by our model.

\par
\medskip
\par
The behavior of the chemical potential  depends on whether there is symmetry between the electron and hole states in the system. If 
the DOS for the electrons and holes is equal or symmetric around the Dirac point, then the chemical potentail does not 
change its sign even for high temperatures, i.e., it remains positive for electron doping or stays negative if $E_F < 0$. 
In fact, these two types of doping result in symmetric behavior, decreasing $\vert \mu(T) \vert$, so that the 
electron-hole symmetry persists at any finite temperature. Let us discuss the electron doping in more detail. Once the 
temperature is sufficiently high, the hole states become thermally excited and it has an opposite effect to that of electron,
decreasing the reduction of the chemical potential. At extremely high temperatures, the two processes have almost equivalent
effects and $\mu(T)$ asymptotically tends to zero and never crosses the Dirac point.
This situation changes if the DOS of the electron and holes differs and the two hole subbands are energetically 
not equivalent, as we observe for TMDC's. We have shown that for MoS$_2$ the chemical potential becomes negative, changing its sign
at $T \backsimeq 2.5 E_F$. Alternatively, $\mu(T)$ could never reach the zero energy line if there is hole doping
at zero temperature and starts decreasing at sufficiently high temperatures.

\par
\medskip
\par
As a necessary intermediate step in our derivations, we obtained a piecewise-linear model for the DOS 
for transition metal dichalcogenides, directly from the Hamiltonian parameters of the considered system. 
This model gives exact results for $\rho_d(\mbb{E})$ at the band edges (next to each energy gap) and a 
fairly good approximation at higher energies. This model significantly improves the results obtained 
from the spin- and valley-resolved gapped graphene approximation.

\par
\medskip
\par
Finally, we considered the way in which the initial doping affects the plasmon dispersions 
$(\hbar \Omega_p/ E_F)^2  \backsimeq \Lambda \, q$, and the effective length $\Lambda$ depends on the doping. 
There could be  an initial carrier density at $T=0$, and the thermally-induced one at  finite temperature. The 
latter doping type results in a finite polarization function and  $\sqrt{q \, T}$ plasmon dispersions even for
intrinsic systems. \,\cite{SDSLi} We demonstrate how much each type of doping contributes this effective length.

\par
\medskip
\par
Several many-body calculations or collective electronic models requires reliable knowledge of 
the chemical potential at finite temperature. In the absence of such information, much attention has
been directed towards intrinsic, or undoped systems, or low temperatures, so that $\backsimeq (T/T_F)^2$ 
series expansions could have been applied. Our results are going to provide considerable assistance   in
transport studies, optical, thermally modulated conductivity for the  materials, discussed in our paper. 
Consequently, we expect our work to provide an important contribution to electronics, 
transport and plasmonics of these recently discovered structures, both theory and experiment.

\acknowledgments
D.H. would like to thank the support from the Air Force 
Office of Scientific Research (AFOSR).

\appendix

\section{Density-of-states}
\label{apa}

The DOS for electrons and holes with energy dispersion $\varepsilon^{\,\xi,\sigma}_{\gamma}(k)$ 
is defined as

\begin{equation}
\rho_d(\mbb{E})= \int \frac{d^2{\bf k}}{(2 \pi)^2} \sum\limits_{\gamma = \pm 1}\,\sum\limits_{{\xi,\sigma} = \pm 1} \,
\delta\left[ \mbb{E} -\varepsilon^{\,\xi,\sigma}_{\gamma}(k) \right] \, ,
\label{ados2}
\end{equation}
where $\xi = \pm 1$ and $\sigma = \pm 1$ are valley and spin indices which in our considerations
always appear as a product $\sigma \xi$, so that a single composite index $\nu = \sigma \xi = \pm 1$ 
could be introduced. This leads to a double degeneracy of all considered dispersion relations and 
a transformation  $\sum\limits_{\xi,\sigma = \pm 1} \Longrightarrow 2
\cdot \sum\limits_{\nu = \pm 1} \, ,
$ which we will use throughout this work.

\par
\medskip
\par

For a number of cases of 2D structures, Eq.\eqref{dos} could be immediately evaluated 
using the following property of a delta function
\begin{equation}
 \delta(f(x)) = \sum\limits_{i} \frac{\delta(x-x_i)}{\vert df(x)/dx_{\,x=x_i} \vert} \, ,
\end{equation}
where $x_i$ are the roots of $f(x)$, formally given as $f(x) = \mbb{E} - \epsilon_\nu^{\gamma}(k)$ for various 
energy dispersions $\epsilon_\nu^{\gamma}(k)$. In the case of silicene and germanene, the result is 
straightforward with

\begin{equation}
 \rho_d(\mbb{E}) = \frac{1}{\pi} \, \sum\limits_{\gamma = \pm 1}  
\frac{\gamma \,\mbb{E}}{\hbar^2 v_F^2} \sum\limits_{i = <,>} \Theta
 \left[ \frac{\mbb{E}}{\gamma} - \Delta_i
 \right] \  , 
\end{equation}
i.e., the DOS for Dirac gapped systems is linear similar to graphene. However, it is finite only above 
the energy gaps. This result also covers the case for gapped graphene  if the two gaps are 
equal $\Delta_< = \Delta_> = \Delta_0$. Furthermore, we arrive at well-known V-shaped 
$\backsim \mbb{E}$ DOS for pristine gapless graphene if $\Delta_{<,>} \rightarrow 0$.  

\par
\medskip
\par

\subsection*{Molybdenum disulfide}

In our work, we discuss several effetive models of different complexity and accuracy, describing the 
energy dispersion $\varepsilon^{\nu}_{\gamma}(k)$ for $MoS_2$. 
First, we consider a spin- and valley- resolved gapped graphene model given by Eq.~\eqref{mo0}, in which
we leave out all the mass $\backsim \alpha$ and $\backsim \beta$ terms. It gives quite accurate results 
for the energy eigenstates next to the corners of Brillouin zone, as shown in Fig.~\ref{FIG:1}.  
However, it is straightforward to see that the corresponding DOS obtained as 

\begin{equation}
 \rho_d(\mbb{E}) = \frac{1}{\pi \, (t_0 a_0)^2} \sum\limits_{\gamma = \pm 1}\, \frac{1}{\gamma}\,\sum\limits_{\nu= \pm 1} \,  \, \left(
 \mbb{E} - \frac{\nu}{2}\lambda_0 
 \right) \, \Theta \left[
\gamma\left(\mbb{E} -\frac{\nu\lambda_0}{2}\right) - \frac{1}{2} \left( \Delta - \nu\lambda_0 \right)
\right] \, ,
\label{gap-a}
\end{equation}     
is V-shaped and does not match 
the numerical results even near the band edges. Once we get into the "allowed`` energy ranges 
outside the band gap  $\varepsilon^{\nu=-1,1}_{\gamma = 1}(k = 0) = \Delta/2$ for electrons,
$\varepsilon^{\nu=-1}_{\gamma = -1}(k = 0) = -\Delta/2 + \lambda_0$ and $\varepsilon^{\nu=1}_{\gamma 
= -1}(k = 0) = -\Delta/2 - \lambda_0$ for the holes, the DOS experiences three different
giant discontinuities due to each new contributing energy subband. These giant leaps could be   
calculated using the parabolic subbands approximation, obtained for $k \ll k_F$ 

\begin{equation}
\varepsilon_{\gamma}^{\nu}(k)  = \frac{1}{2} \left[ 
\nu\lambda_0 (1 - \gamma) + \gamma \, \Delta 
\right] + \left[
\frac{\hbar^2}{4 m_e} (\alpha + \gamma \beta ) + \frac{\gamma \, (t_0 a_0)^2}{\Delta -\nu\lambda_0}
\right] k^2 \, .
\label{a-parab}
\end{equation}
This result leads to the DOS given by \,\cite{ourjap2017}

\begin{equation}
\rho_d(\mbb{E}) = \frac{1}{2\pi\hbar^2} \sum \limits_{\gamma,\,\nu = \pm 1}
\Big| \frac{\alpha+\gamma \beta}{4 m_e} + \frac{\gamma (t_0a_0)^2}{ \hbar^2(\Delta - \nu\lambda_0)} \Big|^{-1}
\Theta \left[
\gamma\left(\mbb{E} -\frac{\nu\lambda_0}{2} \right) - \frac{1}{2} \left( \Delta -\nu\lambda_0 \right)
\right]\, .
\label{a-jumps}
\end{equation}
The two terms in Eq.~\eqref{a-jumps} are of the same order of magnitude, 
consequently, each of them must be retained in our calculation. Physically it means that due to
large gap $\Delta$ the curvature of each subband at $k = 0$ is so small that even the $\backsimeq \alpha/(4 m_e)$ 
correction is significant. It basically discards the DOS obtained for the gapped graphene model,
even at $k \rightarrow 0$. The fact that the mass and even higher order terms must be taken into account 
for the plasmon calculation was mentioned in Ref.~[\onlinecite{Scholz1}].

\par
\medskip
\par

The principal model which yields quite reliable results for the DOS   
is derived by neglecting only non-essential ${\cal O}(k^4)$ terms with

\begin{equation}
\varepsilon^{\nu}_{\gamma}(k)\backsimeq \frac{1}{2}\,\nu \, \lambda_0 + \frac{\hbar^2\,\alpha}{4 m_e} \,k^2 + 
\frac{\gamma}{2}
\sqrt{
\left(\Delta - \nu \, \lambda_0 \right)^2
+
\left[
(2 t_0 a_0)^2+\left( \Delta - \nu \, \lambda_0  \right) \frac{\hbar^2 \, \beta}{m_e} 
\right] k^2 
} \,\,\, .
\label{ksq-a}
\end{equation}
Here, all $\backsim k^2$ terms are retained and the final expression for the DOS appears to be quite complicated. 
We use a general equation from our previous work \,\cite{ourjap2017} to come up with a linear approximation 
which, according to the most precise and generalized numerical results, \,\cite{Scholz1, ourjap2017} is valid 
for all experimentally allowable electron and hole doping densities. For the dispersions Eq.~\eqref{ksq-a}, 
we use Eq.~\eqref{ados2} to obtain

\begin{equation}
\rho_d(\mbb{E}) = \frac{1}{2\pi} \, \sum\limits_{j} \, \sum\limits_{\gamma,\,\nu=\pm 1} 
\Bigg|
\tilde{\alpha} + \frac{\gamma \, \tilde{A}_\nu(\Delta \pm \lambda_0, \beta \, \vert \, a_0 t_0) }{ 2 \left\{
\mbb{E}-\tilde{\epsilon}_\nu -\tilde{\alpha} \, \xi_\nu^{\,(j)} (\mbb{E})
\right\}
} 
\Bigg|^{-1} 
\Theta \left[
\gamma\left(\mbb{E} -\frac{\mu\lambda_0}{2}\right) - \frac{1}{2} \left( \Delta - \nu\lambda_0 \right)
\right]\, ,
\label{rhofull}
\end{equation}
where $\tilde{\epsilon}_\nu=\nu \lambda_0/2$, $\tilde{\Delta}_\nu=(\Delta -\nu\lambda_0)/2$,
$\tilde{A}_\nu (\Delta \pm \lambda_0, \beta \, \vert \, a_0 t_0)=(\Delta -\nu\lambda_0)\hbar^2 \beta /(4 m_e) +
(t_0 a_0)^2$, and $\tilde{\alpha}=\hbar^2\alpha/(4 m_e)$. $\xi_\nu^{\,(j)}(\mbb{E})$ are the roots of 

\begin{equation}
f(\mbb{E},\xi) = \mbb{E} - \tilde{\epsilon}_\nu - \alpha \xi - \gamma \, \sqrt{\tilde{A}_\nu \xi + 
\tilde{\Delta}_\nu^2} = 0 \, .
\label{eq}
\end{equation}
This euqation could be solved by expressing it  in quadratic form. However, one must bear in mind that if both 
parts of Eq.~\eqref{eq} are squared, there might be additional non-physical solutions which must be disregarded.  
If this equation is written as $(\alpha \xi)^2 + B \, \xi + C = 0$, where $B = \tilde{A}_\nu + 2 \, 
\alpha$ and $C = (\mbb{E} - \tilde{\epsilon}_\nu)^2 - \tilde{\Delta}_\nu^2$, the only appropriate 
solution is $ \xi^{\,(1)}(\mbb{E}) = 1/(2 \alpha)^2 \left( B + \sqrt{B^2 - 4 \alpha^2 C}\right)$. 
The other solution
corresponds to the  $\mbb{E} - \tilde{\epsilon}_\nu - \alpha \xi  = - \gamma \, 
\sqrt{\tilde{A}_\nu \xi + \tilde{\Delta}_\nu^2}$ and is obviosly incorrect
for $\alpha \rightarrow 0$.  We also note that the electron/hole index $\gamma$ is no longer present 
in this equation, so that the two $\pm$ solutions are not associated with electron or hole states.

\par
\medskip
\par
In order to illustrate the physics behind selecting the only appropriate solution, let us consider a simple 
example of gapless graphene with $\varepsilon_{\gamma}(k) = \hbar v_F k $ with additional small, not depending on
$\gamma$, the mass term $\backsim \alpha k^2$, $\alpha \ll \hbar v_F/ k_F$. The actual dispersion relation is now 
$\epsilon^{\gamma}(k) = \gamma \hbar v_F \vert k \vert + \alpha k^2$, and the DOS is calculated as

\begin{equation}
 \rho_d(\mbb{E}) = \frac{2}{\pi} \, \sum\limits_{\gamma = \pm 1} \, \sum\limits_{j} \frac{
 k^{\,(j)}}{
 \vert
 \gamma \, \hbar v_F + 2 \alpha k^{\,(j)}
 \vert
 } \, .
\end{equation}
The roots $k^{\,(j)}$ are the solutions of $\gamma \hbar v_F \vert k \vert + \alpha k^2 = 0$. Even though
 such a quadratic equation generally has two inequivalent solutions, only one of them 
$\pm k^{(1)} \backsimeq \mbb{E}/(\gamma \hbar v_F) - \alpha \, \mbb{E}^2/(\hbar^3 v_F^3)$
satisfies the $\vert k \vert$-type equation. Consequently, we obtain the following expression

\begin{equation}
  \rho_d(\mbb{E}) \backsimeq \frac{2}{\pi} \, \sum\limits_{\gamma = \pm 1} \, 
  \frac{\mbb{E}}{\gamma (\hbar v_F)^2} - 
  \frac{ \alpha \,(2 + \gamma) \,\mbb{E}^2}{
   (\hbar v_F)^4} 
\end{equation}
becoming equivalent to graphene DOS $\rho_d{(\mbb{E})} = 2/\pi \, \mbb{E}/ \left[\gamma  (\hbar v_F)^2 \right]$ 
for $\alpha \rightarrow 0$. For the holes with $\gamma = -1$, however, the linear and quadratic mass terms are 
competing, so that for $ k \gg k_F$, another solution is present. However such wave vectors are beyond the Dirac 
cone model, and therefore, it is not physically acceptable. The obtained correction to the DOS is a small decrease, 
as it is expected to be for an energetically elevated location of the subband.
However, in our model for $MoS_2$, the mass terms $\backsim \alpha$ and $\backsim \beta$ are not small and represent a 
finite correction to the DOS. The small parameter which we used in our series expansions is the energy 
$\delta \epsilon$ above each band gap.

\par
\medskip
\par

Now, we return to Eq.~\eqref{eq} and present its solution as

\begin{equation}
 \xi^{\,(1)} = \frac{1}{2 \, \tilde{\alpha}^2} \, \left\{
 \tilde{A}_\nu + 2 \, \tilde{\alpha} \left( \mbb{E} - \tilde{\varepsilon}_\nu^{(0)} \right)
 - \left[
 \tilde{A}_\nu^2 + 4 \, \tilde{\alpha}^2 \tilde{\Delta}_\nu^2 + 4 \, \tilde{\alpha}  \, \tilde{A}_\nu \, \left( \mbb{E} -
 \tilde{\varepsilon}_\nu^{(0)} \right)
 \right]^{1/2}
 \right\} \, .
 \label{ksi1}
\end{equation}
This solution is exact in the sense that no approximations have been made so far excpet for 
the $\backsim k^2$ dispersions \eqref{ksq-a}. Substituted into Eq.~\eqref{rhofull}, it gives 
the DOS for an arbitrary energy, for both electrons and holes.

\par
\medskip
\par

As the next step, we substitute this result for $\xi^{\,(1)}$ into Eq.~\eqref{rhofull}.
We are interested in obtaining a linear approximation of the DOS next to 
each subband edge. Let us first consider electrons with  $\varepsilon^\nu_{1}(k) = 
\Delta/2 + \delta \epsilon$, $\delta \epsilon \ll \mbb{E}$. In this case,

\begin{equation}
 \xi_\nu^{\,(1)} \backsimeq \frac{4 m_e}{\alpha \, \hbar^2} \, \left\{ 
 1 - \frac{
 (a_0 \, t_0)^2 + \hbar^2 \beta/(4 m_e) \, \left( \Delta - \nu \lambda_0 
 \right)
 }{
 (a_0 \, t_0)^2 + \hbar^2 /(4 m_e) \, \left[
 (\alpha + \beta) \left( \Delta - \nu \lambda_0 
 \right)
 \right]
 }
 \right\} \, \delta \epsilon \, ,
\end{equation}
and the DOS is now approximately given by  

\begin{eqnarray}
 && \rho_d (\mbb{E}) = \frac{1}{2 \pi} \, \sum\limits_{\nu=\pm 1} \, \\
 \nonumber
 && \frac{
 \Delta - \nu \lambda_0
 }{
 (a_0 t_0)^2 + \hbar^2/(4 m_e) \,(\alpha + \beta)(\Delta - \nu \lambda_0)
 } + 
 \frac{ 2 \, \delta \epsilon \,\,
 \left[
 (a_0 \, t_0)^2 + \hbar^2 \beta/(4 m_e) \, \left( \Delta - \nu \lambda_0 
 \right) \,
 \right]^2
 }{
 \left\{ 
 (a_0 \, t_0)^2 + \hbar^2 /(4 m_e) \, \left[
 (\alpha + \beta) \left( \Delta - \nu \lambda_0 
 \right)
 \right] \,
 \right\}^3
 }  \, . 
\end{eqnarray}
The actual numerical resuts are  determined from
  
\begin{eqnarray}
\nonumber
&& \rho_d (\mbb{E}) = c_0^{(3)} + c_1^{(3)} \left( \mbb{E} - \frac{\Delta}{2} \right) \, , \\
\nonumber 
&& c_0^{(3)} = 0.180\, \frac{1}{t_0 a_0^2} = 11.74 \, \frac{E_F^{(0)}}{(\hbar v_F)^2} \, , \\
&& c_1^{(3)} = 0.268 \, \frac{1}{(t_0 a_0)^2} = 1.218 \, \frac{1}{(\hbar v_F)^2} \, .
\label{c3}
\end{eqnarray}
In the valence band, we consider two separate hole subbands with $\nu = \pm 1$. 
If $\gamma = -1$, $\nu = 1$ and $\mbb{E} \leq -\Delta/2 + \lambda_0$,
the DOS is $\rho_d(\mbb{E}) = c_0^{(2)} + c_1^{(2)} \, \left[ \mbb{E} -
\left( \frac{\Delta}{2} - \lambda_0  \right) \right]$, and the expansion coefficients are

\begin{eqnarray}
 \label{c2a}
 && c_0^{(2)} = \frac{1}{2 \pi} \, \frac{\Delta - \lambda_0}{(a_0 t_0)^2 + (\beta - \alpha) (\Delta - \lambda_0)}  \, , \\
 \nonumber
 && c_1^{(2)} = \frac{1}{\pi} \, \frac{ \delta \epsilon \,\,
 \left[
 (a_0 \, t_0)^2 + \hbar^2 \beta/(4 m_e) \, \left( \Delta - \lambda_0 
 \right) \,
 \right]^2
 }{
 \left\{ 
 (a_0 \, t_0)^2 + \hbar^2 /(4 m_e) \, \left[
 (\beta - \alpha) \left( \Delta - \lambda_0 
 \right)
 \right] \,
 \right\}^3
 } < 0 \, ,
\end{eqnarray}
or

\begin{eqnarray}
\nonumber
 && c_0^{(2)} = 0.105\, \frac{1}{t_0 a_0^2} = 6.847 \, \frac{E_F^{(0)}}{(\hbar v_F)^2} \, , \\
 && c_1^{(2)} = -0.232 \, \frac{1}{(t_0 a_0)^2} = - 1.051 \, \frac{1}{(\hbar v_F)^2} \, .
 \label{c2}
 \end{eqnarray}

\par
\medskip
\par

Finally, for the lower hole subband with $\mbb{E} \lessapprox - \Delta/2 - \lambda_0$, 
we obtain

\begin{eqnarray}
 &&  \rho_d(\mbb{E}) = c_0^{(2)} + c_1^{(2)} \, \left[ \mbb{E} - \left( \frac{\Delta}{2} + \lambda_0  \right) \right] \, , \\
 \nonumber
 && c_0^{(1)} = \frac{1}{2 \pi} \, \sum\limits_{\nu = \pm 1} \frac{\Delta -\nu \, \lambda_0}{(a_0 t_0)^2 + (\beta -
 \alpha) (\Delta - \nu \, \lambda_0)} 
 \, , \\
 \nonumber
 && c_1^{(1)} = \frac{1}{\pi} \, \, \sum\limits_{\nu = \pm 1} \frac{ \delta \epsilon \,\,
 \left[
 (a_0 \, t_0)^2 + \hbar^2 \beta/(4 m_e) \, \left( \Delta -\nu \, \lambda_0 
 \right) \,
 \right]^2
 }{
 \left\{ 
 (a_0 \, t_0)^2 + \hbar^2 /(4 m_e) \, \left[
 (\beta - \alpha) \left( \Delta - \nu \, \lambda_0 
 \right)
 \right] \,
 \right\}^3
 } < 0 \, , 
\end{eqnarray}
and 

\begin{eqnarray}
\nonumber
 && c_0^{(1)} = 0.233\, \frac{1}{t_0 a_0^2} = 15.17 \, \frac{E_F^{(0)}}{(\hbar v_F)^2} \, , \\
 && c_1^{(1)} = -0.458 \, \frac{1}{(t_0 a_0)^2} = 2.077 \, \frac{1}{(\hbar v_F)^2} \, .
 \label{c1}
\end{eqnarray}
 It is straightforward to obtain our previous results for the gapped graphene model \eqref{gap-a} DOS 
if $\alpha = \beta \rightarrow 0$. We also note that the slope of the DOS in the conduction 
band is negative, as it should be accoring to Fig.~\ref{FIG:1} $(d)$, and the summation over the
$\nu$ index is present in all cases excpet the upper hole subband in Eqs.~\eqref{c2a}.

\par
\medskip
\par 

Our results  \eqref{c3} - \eqref{c1} (here, we move from the conduction electrons to the valence band,
i.e., from the right to left) represent a fairly good match with the previously obtained numerical
values, specified in Sec.~\ref{s2} and later used for all our finite temperature calculations. The 
coefficients $c_0^{(i)}$, $i = 1, 2, 3$ are equal to the giant discontinuities of the DOS
$\delta \rho_d (\mbb{E})$ at each subband edge or $k = 0$, except $c_0^{(1)} \backsimeq \delta \rho_d^{(2)}
(-\Delta/2 + \lambda_0) + \delta \rho_d^{(1)} (-\Delta/2 - \lambda_0)$), and, therefore, are accurate.
The linear coefficients $c_1^{(i)}$ , in fair agreement, are $20-25 \%$ larger compared with the 
numerical results since all the $\backsim k^4$ terms of our energy dispersions are neglected. Inclusion 
of these terms leads to the higher energies for chosen wave vector and a decrease of the     
DOS. This discrepancy is increased for higher energies, which is well seen for $c_1^{(1)}$ for holes with 
$\mbb{E} < -\Delta/2 - \lambda_0$. However,  the subbands for such energy ranges do not receive any substantial 
doping unless the temperature becomes very high. For most considered situations, we are limited for $\delta \epsilon
\approx \lambda_0$ within the band edges. In such a small range, $c_1^{(i)}\, \delta \epsilon  \ll c_0^{(i)} $, so 
that the actual DOS values remain almost unaffected and our model yields accurate results.

\section{Chemical potential $\mu=\mu(T)$ at a finite temperature }
\label{apb}

We now derive a set of algebraic equations for the finite-temperature chemical potential
$\mu(T)$. At zero temperature, it is equal to the Fermi energy $E_F = \mu(T)\big\vert_{T=0}$. Our derivation
is based on the total carrier density conservation, which includes both electrons and holes, for zero and any
finite temperatures \,\cite{SDSS}

\begin{equation}
 n = n^{(e)} + (-1) n^{(h)} = \int\limits_0^{\infty} d \mbb{E} \, \rho_d(\mbb{E}) f_{\gamma = 1}(\mbb{E},T) - 
 \int\limits_{-\infty}^{0} d \mbb{E} \, 
\rho_d( \mbb{E} )  \left\{ 1 - f_{\gamma = 1}(\mbb{E,T}) \right\} \, .
\label{ECons}
\end{equation}
The electron and hole occupation probabilities are complimentary and for electron
doping at $T = 0$ the hole states term has no effect on Eq.~\eqref{ECons} (for details see 
Ref.~[\onlinecite{proj}]).

\par
\medskip 
\par

We begin  with the relatively simple case for silicene with dispersions \eqref{sildisp}. The DOS 
$\rho_d(\mbb{E})$ for which buckled honeycomb lattices is given by Eq.~\eqref{dos}. The expression for the 
Fermi energy $E_F$ for fixed electron doping density $n$ at zero temperature depends on whether either one or 
both electron subbands are doped. The former case occurs for doping densities 

\begin{equation}
n \leq n_c = \frac{1}{2 \pi} \, \frac{\Delta_>^2 - \Delta_<^2}{\hbar^2 v_F^2} = \frac{2}{\pi \hbar^2 v_F^2} \, \Delta_{SO} \Delta_z \, ,
\label{nc}
\end{equation}
and the Fermi energy is obtained form 

\begin{equation}
 n = \frac{1}{2 \pi} \, \frac{E_F^2 - \Delta_<^2}{\hbar^2 v_F^2} \, .
 \label{n1}
\end{equation}
Alternatively, if the doping density is sufficient to populate both subbands, $E_F$ is determined by
\begin{equation}
n = \frac{1}{\pi} \, \frac{1}{\hbar^2 v_F^2} \, \left[ E_F^2 - \frac{1}{2} \, \left( \Delta_<^2 + \Delta_>^2\right) \right] \, .
\label{n2}
\end{equation}

\par
\medskip
\par

Once the temperature is set finite, Eq.~\eqref{ECons} leads us to  $n = \mc{I}^{(e)}(\Delta_i,T)
-\mc{I}^{(h)}(\Delta_i,T)$, with the two terms corresponding to electron and hole components of 
the total carrier density. These integrals are presented as 

\begin{equation}
\mc{I}^{(e)}(\Delta_i,T)  =  \int\limits_{\Delta_<}^{\infty} d \mbb{E} \, \mc{A} (\mbb{E},T) + 
  \int\limits_{\Delta_>}^{\infty} d \mbb{E}  \, \mc{A} (\mbb{E},T) \, , 
\label{e1}
\end{equation}
where

\begin{equation}
 \mc{A}(\mbb{E},T) = \frac{1}{\pi} \, \frac{\mbb{E}}{\hbar^2 v_F^2} \, 
\left\{1  +  \tet{exp} \left[ \frac{\mbb{E} - \mu(T)}{k_B T} \right] \right\}^{-1} \, .
\end{equation}
Each of the these integrals could be easily evaluated. Using a variable substitution
 $\xi = (\mbb{E} - \Delta_< )/k_B T$,   we obtain 

\begin{equation}
\frac{1}{\pi} \, \frac{1}{\left(\hbar v_F \right)^2} \int\limits_{\Delta_<}^{\infty} d \mbb{E} \, \mbb{E} \, 
\left\{1  +  \tet{exp}
\left[ \frac{\mbb{E} - \mu(T)}{k_B T} \right] \right\}^{-1}
= \frac{k_B T}{\pi \left(\hbar v_F \right)^2} \int\limits_{0}^{\infty} d \xi \, (\Delta_< + \xi \, k_B T) \, 
\left\{ 1 + \tet{exp} \left[\xi - \frac{\mu(T) - \Delta_<}{k_B T} \right] \right\}^{-1}
\end{equation}
With the help of the following notation

\begin{equation}
 \mc{R}^{(p)} (T,X) = \int\limits_{0}^{\infty} d \xi \,\, \xi^p/ \left\{1 + \tet{exp}[\xi - X / (k_B T) ] \right\} \, ,
 \label{R}
\end{equation}
we obtain the final result of the integration as

\begin{equation}
\frac{k_B T \, \Delta_<}{\pi \left( \hbar v_F\right)^2} \, \mc{R}^{(0)} [T, \, \mu(T) - \Delta_<] +
  \frac{1}{\pi} \, \left(\frac{k_B T}{\hbar v_F} \right)^2 \, \mc{R}^{(1)} [T, \, \mu(T)-\Delta_< ] \, .
\end{equation}
For $p=0$ and $1$, corresponding to the  2DEG and gapless graphene, Eq.~\eqref{R} leads to \,\cite{wcrc, math2}

\begin{eqnarray}
\nonumber
&& \mc{R}^{(0)} (T, X) = \ln \left\{ 1 + \tet{exp} \left[\frac{X}{k_B T}\right] \right\} \, , \\
&& \mc{R}^{(1)} (T, X) = - \text{Li}_{\,2} \left\{ - \tet{exp} \left[\frac{X}{k_B T} \right] \right\} \, ,
\end{eqnarray}
where $\text{Li}_{\,2} (z)$ is the \textit{second-order polylogarithm function} or 
\textit{dilogarithm} defined as

\begin{eqnarray} 
  \nonumber
 && \text{Li}_{\,p} (z) = \sum\limits_{k=1}^{\infty} \frac{z^k}{k^p} \, , \\ 
 && \text{Li}_{\,2} (z) = - \int\limits_0^{z} \frac{\ln (1-t)}{t} \, dt \, .
\end{eqnarray}
The second term of Eq.\eqref{e1}, which only differs from the first one by its integration limits, is 

\begin{equation}
\frac{2 k_B T \, \Delta_>}{\pi \left( \hbar v_F \right)^2} \, \ln \left\{ 1 + 
\tet{exp} \left[\frac{\mu(T) - \Delta_>}{k_B T}\right] \right\} - \frac{1}{\pi} \, \left( \frac{k_B T}{\hbar v_F} \right)^2 \,
\text{Li}_{\,2} \left\{ - \tet{exp} \left[\frac{\mu(T)-\Delta_>}{k_B T} \right] \right\} \, .
\end{equation}
The remaining term $\mc{I}^{(h)}(\Delta_{i},T)$, $i = \{ <,> \}$, which describes the contribution from the holes, 
is also easily obtained

\begin{eqnarray}
 \nonumber
  && \mc{I}^{(h)}(\Delta_{i},T) = \mc{I}^{(h)}_{1} (\Delta_{i},T) + \mc{I}^{(h)}_2 (\Delta_{i},T) \hskip0.1in \text{where} , \\
 \nonumber 
  && \mc{I}^{(h)}_{1} (\Delta_{i},T) = \frac{1}{\pi} \, \frac{k_B T }{\left( \hbar v_F \right)^2} \,
  \sum\limits_{i = <,>} \Delta_i \, \mc{R}^{(0)} \left\{ T, \, - \left[\mu(T) + \Delta_i \right] \right\} \hskip0.1in \text{and} \\
  && \mc{I}^{(h)}_{2} (\Delta_{i},T) = \frac{1}{\pi} \, \, \left( \frac{k_B T}{\hbar v_F} \right)^2 \,
  \sum\limits_{i = <,>}  \mc{R}^{(1)} \left\{ T, \, - \left[\mu(T) + \Delta_i \right] \right\} \, .
\end{eqnarray}
Now, the total carrier denisty from Eq.~\eqref{ECons} could be written as

\begin{equation}
 n  = \left( \frac{k_B T}{\hbar v_F} \right)^2 \sum\limits_{\gamma = \pm 1} \, \frac{\gamma}{\pi} \, \sum\limits_{i = <,>}
   \mc{R}^{(1)} \left[ T, \, \gamma \mu(T) - \Delta_i  \right] +
  \frac{\Delta_i}{k_B T} \, \mc{R}^{(0)} \left[ T, \, \gamma \mu(T) - \Delta_i \right] \,
\end{equation}
or, explicitly expressing the polylogarithm functions, we write

\begin{equation}
n \, \left( \frac{\hbar v_F}{k_B T} \right)^2 =  \sum\limits_{\gamma = \pm 1} \, \frac{\gamma}{\pi} \, \sum\limits_{i = <,>}
- \text{Li}_{\,2} \left\{ - \tet{exp} \left[\frac{ \gamma \mu(T) - \Delta_i}{k_B T} \right] \right\}  +
\frac{\Delta_i}{k_B T} \,  \ln \left\{ 1 + \tet{exp} \left[\frac{\gamma \mu(T) - \Delta_i}{k_B T}\right] 
\, \right\} \, .
\label{musil}
\end{equation}

Using Eqs.~\eqref{n1} or Eq.~\eqref{n2} depending on whether only one or both  subbands are filled at zero temperature, 
we obtain the equation which relates the finite-temperature chemical potential with its $T = 0$ value $E_F$.   
Energy band gap(s) obviously affects this result. 

\par
\medskip
\par

The $\mu(T)$ for gapped graphene with two fourfold degenerate energy subbands is obtained if we substitute  
$\Delta_< = \Delta_> = \Delta_0$ and $\sum\limits_{i = <,>} \Longrightarrow  \times 2$.
For gapless graphene, $\Delta_0 = 0$ and $\pi n = \left[ E_F/ (\hbar v_F) \right]^2$, so that we write

\begin{equation}
 \frac{1}{2 \left( k_B T \right)^2} \, E_F^2  =  \sum\limits_{\gamma = \pm 1} \gamma \, \mc{R}^{(1)}[T, \, \gamma \mu(T) ]
 = - \sum\limits_{\gamma = \pm 1} \gamma \,\text{Li}_{\,2} \left\{ - \tet{exp} \left[\frac{ \gamma \, \mu(T)}{k_B T} \right] \right\} 
 \, .
\end{equation}
If the temperature is kept low with $k_B T \ll E_F$, this result is simplified as \,\cite{SDSS}

\begin{equation}
\mc{R}^1 (x) \backsimeq \left( \frac{x^2}{2}+ \frac{\pi^2}{6} \right) \, \Theta(x) + x \,
\ln \left(1 + \tet{e}^{- \vert x \vert} \right) \, ,
\end{equation}

\par
\medskip
\par

Finalizing our derivations for silicene, we briefly address the case of hole doping with $E_F < 0$. The left 
part of Eq.~\eqref{ECons} is now modified as

\begin{equation}
 - n^{(h)} = - \frac{2}{\pi} \frac{1}{(\hbar v_F)^2} \sum\limits_{i = < ,>} \, \int\limits_{- \infty}^{-\Delta_i}
 d \,\mbb{E} \, \vert \mbb{E} \vert \,\, \Theta(-\mbb{E} + E_F) \, . 
\end{equation}
In analogy with electron doing, the Fermi energy depends on whether only the $\Delta_<$-subband (which 
is now the higher one) or they are both  doped. The equations determining the Fermi energy for given hole 
doping density $n$ are exactly similar to Eqs.~\eqref{n1} and \eqref{n2}, which confirms complete symmetry 
between the electron and hole states in silicene. The right part of Eq.~\eqref{ECons} remains unchanged 
except the chemical potential is negative $\mu < 0$ for any finite temperature.

\subsection*{$\mu(T)$ for transition metal dichalcogenides} 

In cotrast to the previously considered buckled honeycomb lattices and graphene, the electron/hole symmetry 
in TMDC's (such as $MoS_2$) is clearly broken. Even in the simplest gapped graphene model given by 
Eq.~\eqref{mo0}, the two hole subbands are not degenerate and separated by $\lambda_0$ at $k=0$. For all 
reasonable doping densities $n < 10^{17}\,m^{-2}$, the Fermi energy is such that the lower $\varepsilon_{\nu 
= 1}^{\gamma = -1}(k=0)= -\Delta/2 - \lambda_0$ subband \textit{could not be populated} at zero temperature. 
This could be verified by rewriting Eq.~\eqref{nc} as

\begin{equation}
n_c = \frac{2}{\pi} \frac{\lambda_0 \, \Delta}{\left(t_0 \, a_0\right)^2}  = 1.0 \cdot 10^{18}\,m^{-2}\, .
\label{nc0Mo}
\end{equation}

\par
\medskip
\par

From here on, we are going to use a picewise-linear model for the DOS with the 
empirical coefficients, provided in Sec.~\ref{s2}. Our analytical model for the DOS, 
develop in the preceding Appendix in Eqs.~\eqref{c3} - \eqref{c1} could also be employed here without
losing much of precision.   Let's us first consider electron doping with density $n_{(0)}^{e}$ at zero temperature.  
The corresponding Fermi energy is determined by

\begin{equation}
 n_{(0)}^{e} = c_1^{(3)}/2 \, \left( E_F^2 - \Delta^2/4 \right) + c_0^{(3)} \, \left( E_F - \Delta/2 \right) \, , 
 \label{nemo}
\end{equation}
or 

\begin{equation}
 E_F^{\,e} = \frac{1}{c_1^{(3)}} \, \left\{- c_0^{(3)} + \left[ \left( c_0^{(3)} + c_1^{(3)} \, 
\frac{\Delta}{2} \right)^2 + 2 n_{(0)}^{e} 
 c_1^{(3)} \right]^{1/2} \right\} \, .
\end{equation}
For hole doping the result is quite similar, except $c_1^{(2)} < 0$  and the upper valence 
band gap is $-\Delta/2 + \lambda_0$:

\begin{equation}
 E_F^{(h)} =  \frac{1}{c_1^{(2)}} \, \left\{ 
 - c_0^{(2)} + \left\{ 
 -2 c_1^{(2)} \, n_{(0)}^{h} + \left[
 c_0^{(2)} - c_1^{(2)} \,
 \left(
 \frac{\Delta}{2} - \lambda_0
 \right)
 \right]^2 
 \right\}^{1/2} \,
 \right\} \, .
\end{equation}
Using this expression, we can further improve the result in Eq.~\eqref{nc0Mo} for the critical hole doping
density $n_c$ needed to reach the lower subband at zero temperature. The required Fermi energy is $E_F^{(h)} 
\leq - \Delta/2 - \lambda_0$, so that the corresponding hole density is  
$ n_{c}^{h} = - \lambda_0 \, \Delta \,\, c_1^{(2)} + 2 c_0^{(2)} \, \lambda_0 = 3.3 \cdot 10^{17}\, m^{-2} $.

This critical density value is still far above the experimentally realizable values $ \backsimeq 1.0 \cdot 10^{17}\, 
m^{-2}$, and for all our calculations, the lower hole subband \textit{is never populated} at $T=0$.

\par
\medskip
\par

Finally, we address the finite-temperature chemical potential for $MoS_2$. Once again, we start with the carrier 
density conservation equation \eqref{ECons}.


The electron component is easily evaluated as


\begin{equation}
k_B T \, \left(c_0^{(3)} + \frac{\Delta}{2} \right) \, \mc{R}^{(0)} \left[T, \, \mu(T) - \frac{\Delta}{2} \right] +
c_1^{(3)} \left(k_B T\right)^2 \, \mc{R}^{(1)} \left[T, \, \mu(T)-\frac{\Delta}{2} \right] \, .
\label{Iemo}
\end{equation}
 
\par
 \medskip
 \par
The hole state integrals are written in the following form

\begin{eqnarray}
&& \int\limits_{-\infty}^{0} d \mbb{E} \,  \rho_d( \vert \mbb{E} \vert)  \left\{1 - f_{1}(\mbb{E,T}) \right\}  = 
\int\limits_{-\infty}^{-\Delta/2 + \lambda_0} d \mbb{E} \, \left[ - c_1^{(2)} \mbb{E} + c_0^{(2)} \right]
\left\{1  +  \tet{exp} \left[ \frac{ \mu(T) - \mbb{E}}{k_B T} \right] \right\}^{-1} + \\
\nonumber
&& + \int\limits_{-\infty}^{-\Delta/2 - \lambda_0} d \mbb{E} \, \left[ - \delta c_1^{(1)} \mbb{E} + \delta c_0^{(1)} \right] 
\left\{1  +  \tet{exp} \left[ \frac{\mu(T) - \mbb{E}}{k_B T} \right] \right\}^{-1} \, ,
\end{eqnarray}
where $\delta c_1^{(i)} = c_1^{(i)} - c_2^{(i)}, \, i = 1,2$. They are evaluated as

\begin{eqnarray}
 \nonumber
  && \mc{I}^{(h)}(\Delta,\lambda_0 \, \vert \, T) = \sum\limits_{j = 1} ^{4} \mc{I}^{(h)}_{j} \hskip0.1in \text{where} , \\
 \nonumber 
  &&  \mc{I}^{(h)}_{1} (\Delta,\lambda_0 \, \vert \, T) = k_B T \left(
  \frac{\Delta}{2} - \lambda_0 + c_2^{(0)}
  \right) \,
  \mc{R}^{(0)} \left\{ T, \, - \left[\mu(T) + \frac{\Delta}{2} - \lambda_0 \right] \right\} \, ,\\  
  && \mc{I}^{(h)}_{2} (\Delta,\lambda_0 \, \vert \, T) = c_2^{(1)} \left( k_B T \right)^2 \, 
  \mc{R}^{(1)} \left\{ T, \, - \left[\mu(T) + \frac{\Delta}{2} - \lambda_0 \right] \right\}  \, ,\\
  \nonumber
  &&  \mc{I}^{(h)}_{3} (\Delta,\lambda_0 \, \vert \, T) = k_B T \left(
  \frac{\Delta}{2} + \lambda_0 + \delta c_1^{(0)}
  \right) \,
  \mc{R}^{(0)} \left\{ T, \, - \left[\mu(T) + \frac{\Delta}{2} + \lambda_0 \right] \right\} \, , \\  
  \nonumber
  && \mc{I}^{(h)}_{2} (\Delta,\lambda_0 \, \vert \, T) = \delta c_1^{(1)} \left( k_B T \right)^2 \, 
  \mc{R}^{(1)} \left\{ T, \, - \left[\mu(T) + \frac{\Delta}{2} + \lambda_0 \right] \right\} \, . 
  \label{Ihmo}
\end{eqnarray}
Combined with the electron terms \eqref{Iemo}, these hole integrals \eqref{Ihmo} form the right side 
of the carrier density conservation \eqref{ECons}. Its left side, corresponding to zero temperature,
is given by Eq.~\eqref{nemo} for electron doping and by 

\begin{equation}
 n_{(0)}^{e} = \left( 
 \frac{\Delta}{2} + E_F - \lambda_0
 \right) 
 \left\{
 c_2^{(0)} + \frac{c_2^{(1)}}{2} \left[
 \frac{\Delta}{2} - \left(
 E_F + \lambda_0
 \right)
 \right]
 \right\} \, , 
\end{equation}
for hole doping. The symmetry between the electron and hole states is no longer present, which strongly
affects the finite-temperature behavior of the chemical potential in TMDC's. 

\bibliography{BExFT}

\end{document}